\documentclass{aims} 
\usepackage{amsmath}
\usepackage{paralist}
\usepackage[misc]{ifsym}
\usepackage{epsfig} 
\usepackage{epstopdf} 

\usepackage[colorlinks=true]{hyperref}
\hypersetup{urlcolor=blue, citecolor=red}
\allowdisplaybreaks

\def\currentvolume{X}
 \def\currentissue{X}
  \def\currentyear{200X}
   \def\currentmonth{XX}
    \def\ppages{X-XX}
     \def\DOI{10.3934/xx.xxxxxxx}

\usepackage{algpseudocode}
\usepackage{algorithm}
\usepackage{mathtools}
\usepackage[labelformat=parens,labelsep=space]{subcaption}
\usepackage{array} 
\usepackage{hyperref}

\newtheorem{theorem}{Theorem}[section]

\theoremstyle{definition}
\newtheorem{definition}[theorem]{Definition}
\newtheorem{remark}[theorem]{Remark}
\newtheorem{assumption}[theorem]{Assumption}

\title[Confidence regions for persistence diagrams]
{Confidence regions for a persistence diagram of a single image with one or more loops} 

\author[Susan Glenn Jessi Cisewski-Kehe Jun Zhu and William M Bement]{}


\thanks{Supported by NSF under Grant Number DMS 2038556, DMS 2337243, DMS 2245906; LANL under project number 20240479CR-IST; NIH under Grant Number RO1 GM052932.}

\thanks{$^*$Corresponding author: Susan Glenn}

\begin{document}
\maketitle

\centerline{\scshape
Susan Glenn$^{{\href{mailto:ssglenn@lanl.gov}{\textrm{\Letter}}}*1,2}$, Jessi Cisewski-Kehe$^1$,  Jun Zhu$^1$,  William M Bement$^3$}

\medskip

{\footnotesize
 \centerline{$^1$Department of Statistics, University of Wisconsin, USA}
} 

\medskip

{\footnotesize
 \centerline{$^2$ Los Alamos National Laboratory,
Los Alamos, USA}
}

\medskip

{\footnotesize
 \centerline{$^3$ Center for Quantitative Cell Imaging, 
Department of Integrative Biology, 
University of Wisconsin, USA}
}

\bigskip



{\bf This article has been published in a revised form in Foundations of Data Science  [https://doi.org/10.3934/fods.2025017]. This version is free to download for private research and study only. Not for redistribution, re-sale or use in derivative works.}

\begin{abstract}
Topological data analysis (TDA) uses persistent homology to quantify loops and higher-dimensional holes in data, making it particularly relevant for examining the characteristics of images of cells in the field of cell biology. In the context of a cell injury, as time progresses, a wound in the form of a ring emerges in the cell image and then gradually vanishes. Performing statistical inference on this ring-like pattern in a single image is challenging due to the absence of repeated samples. In this paper, we develop a novel framework leveraging TDA to estimate underlying structures within an individual image and quantify associated uncertainties through confidence regions. Our proposed method partitions the image into the background and the damaged cell regions. Then, pixels within the affected cell region are used to establish confidence regions in the space of persistence diagrams (topological summary statistics). The proposed method establishes an estimate of a persistence diagram for an image that mitigates the bias of a traditional persistence diagram computation. A simulation study is conducted to evaluate the coverage probabilities of the proposed confidence regions in comparison to an alternative approach that is proposed in this paper. We also illustrate our methodology by a real-world example provided by biological cell repair.
\end{abstract}



\begin{figure}
\centering
  \begin{subfigure}{.4\linewidth}
        \includegraphics[width=\linewidth]{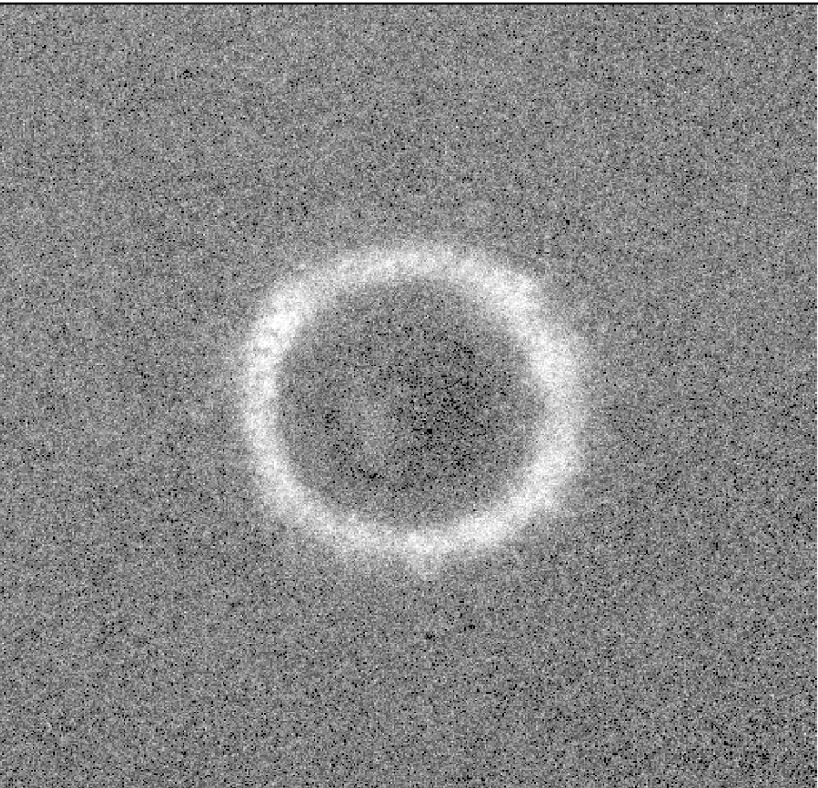}\caption{Control cell wound ring} \label{subfig:con}
  \end{subfigure}  
      \begin{subfigure}{.4\linewidth}
    \centering
        \includegraphics[width=\linewidth]{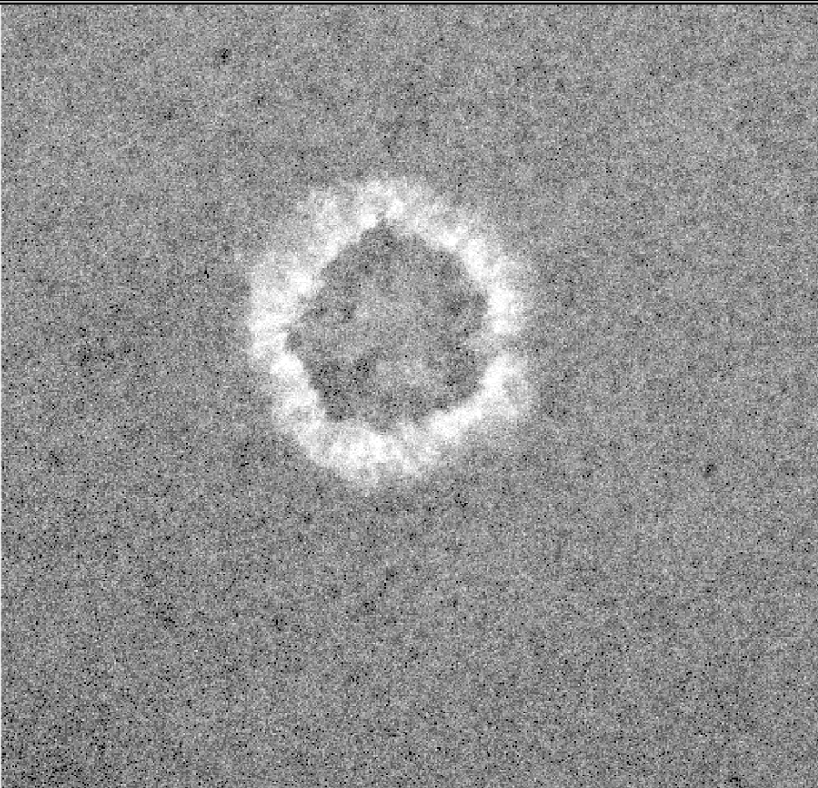}
         \caption{C3 cell wound ring} \label{subfig:c3}
  \end{subfigure} 
  \begin{subfigure}{.1\linewidth}
    \centering
        \includegraphics[width=\linewidth]{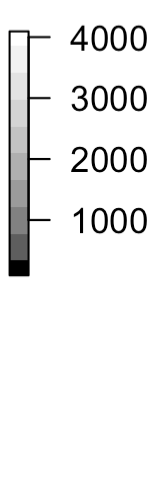}
  \end{subfigure} 
\caption{(A) Grey scale image of a protein ring formed around a wound where there was no toxin (control) injected into the cell. (B) Grey scale image of a protein ring formed around a wound where a toxin (C3) was injected into the cell. }\label{fig:ExampleIntrod}
\end{figure}

Ring-like patterns are ubiquitous in biology, being evident during cell division \cite{1}, development \cite{2}, and the response of immune cells to challenges \cite{3}, to name a few examples. Further, it is not uncommon for ring-like, biological patterns to be perturbed as a consequence of pathological insults \cite{29}. Of particular interest here are the rings of proteins that form around wounds made in single cells as part of the healing response \cite{4}; an example of these patterns can be seen in Figure~\ref{fig:ExampleIntrod}. Such rings close over the wound site, healing the damage, and manipulations that disrupt healing typically alter the organization of the rings \cite{5}. Currently, assessments of wound ring disorganization are largely subjective, or are based on simple comparisons of features like aspect ratios, rather than any metric of underlying ring pattern quality. The purpose of this paper is to develop a statistical method to objectively identify rings and quantify their associated uncertainty.

Topological data analysis (TDA) provides a framework for the quantification of the shape of data. For the wounded cell example, TDA can quantify the pattern of an image by representing each detected ring as a loop on a topological summary statistic called a {\em persistence diagram} introduced in Section~\ref{sec:background}. However, direct inference on persistence diagrams is challenging due to their complex multivariate, multidimensional structure, where even averages are not necessarily unique \cite{6, 7}.

TDA has been applied to analyze a wide range of image processing problems. Much of the current literature is dedicated to machine learning tasks, such as classification or prediction, typically involving multiple images (e.g., \cite{8, 9, 10}). Applications of TDA for inference in image analysis typically involve either multiple images of a single subject or comparisons between two distinct groups (e.g., \cite{11, 12, 8}). When a single image is examined, the focus is often on extracting topological features without addressing statistical inference (e.g., \cite{8, 13}). Notably, there is an existing method for inference on topological features extracted from point cloud data \cite{14}. Overall, there is a dearth in TDA methodology in the context of an image of a ring in a living system, as many existing methods are either designed for point clouds, multiple images, or perform tasks other than inference.

In this paper, we develop a new method for constructing confidence regions for the persistence diagram of a single image. Our focus is specifically on persistence diagrams due to their capability to discriminate and perform inference on individual topological features. The proposed method uses segmentation, dividing the image into contiguous regions, which are subsequently matched to corresponding loops identified in the persistence diagram. These matched loops serve as the basis for estimating the shapes within the underlying pattern such as rings in the case of the current application, though the approach is adaptable to other contexts. The confidence regions built for each matched loop are derived by analyzing the pixel distribution within each partition. This method provides unbiased estimates and asymptotic confidence regions with accurate coverage probabilities. Our proposed method allows for inference on the persistence diagram of a single image which yields a simple intuitive interpretation. While motivated by the wounded cell application, this proposed method generalizes to settings with a single image characterized by one or more loops. In order to have a comparison, we also extend the method in \cite{14} from point clouds to images. 

The remainder of the paper is organized as follows. In Section~\ref{sec:background}, we provide background on TDA and explain how TDA can be applied to analyze the shape of images. In Section~\ref{sec:method}, we present the new method for constructing confidence regions for a persistence diagram of a single image along with an extension of \cite{14} to an image. In Section~\ref{sec:simulations}, both of these methods are used in a simulation study to assess the coverage of the confidence regions of the holes on the persistence diagram. In Section~\ref{sec:CellImages}, we apply our new method to the wounded cell example. We provide conclusions and discussion in Section~\ref{sec:conc}.

\section{Topological data analysis and persistence diagrams} 
\label{sec:background}
This section introduces key principles used in TDA and their application to data in the context of images. First, concepts in algebraic topology, such as persistent homology, are described. Then the focus is on how to characterize the intrinsic shape and structure of an image and represent this information on a persistence diagram.

TDA uses ideas from algebraic topology and computational geometry to extract meaningful insights and patterns from data. In particular, persistent homology is used to quantify the shape of a dataset through identifying holes in the space and determining their number, strength (through persistence), and dimension. 

Homology associates algebraic structures, called homology groups, with topological spaces. These groups $H_p(X)$, where $p$ represents the homology group dimension,  can be thought of as characterizing a topological space $X$ by the number of connected components (the number of zero-dimensional homology group generators, $H_0(X)$) and the number of loops (the number of one-dimensional homology group generators, $H_1(X)$) in $X$  \cite{15, 16}. When $p \geq 2$, $H_p(X)$ correspond to higher dimensional holes in $X$. In this paper, we restrict the focus to the first homology group ($H_1$) since the interest is in the loops, or rings, in an image. Figure~\ref{fig:ExampleIntrod} presents two wounded cells (a control wound and a toxin-exposed wound), both of which exhibit ring-like wound patterns. Differences in wound healing can reveal underlying important properties in the biological responses, which may not be visually obvious as illustrated by the apparent similarity between the two cells displayed in this figure. Persistent homology tracks the evolution of these homology groups across various scales \cite{17, 16}. 

\begin{figure}
\centering
  \begin{subfigure}{.37\linewidth}
        \includegraphics[width=\linewidth]{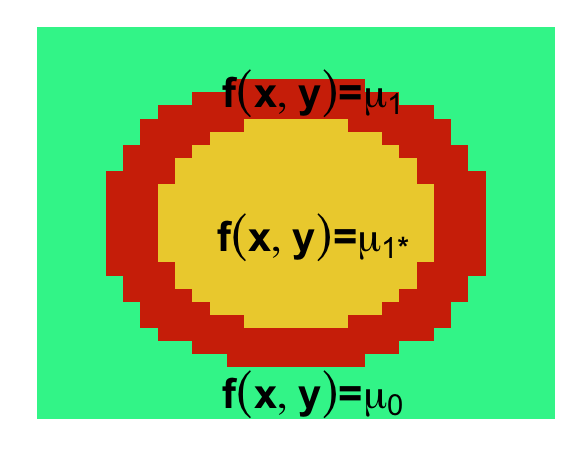}
         \caption{Pattern $\mathcal{M}^0$} \label{subfig:Pattern}
  \end{subfigure}  
      \begin{subfigure}{.37\linewidth}
    \centering
        \includegraphics[width=\linewidth]{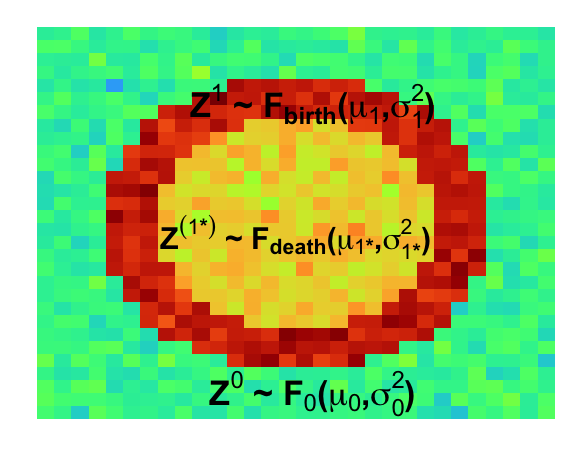}
         \caption{Observed $\mathcal{M}^{\sigma}$} \label{subfig:Observered}
  \end{subfigure} 
        \begin{subfigure}{.1\linewidth}
        \includegraphics[width=\linewidth]{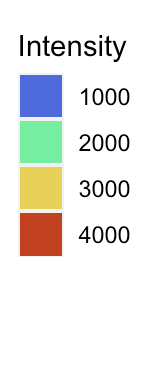}
  \end{subfigure} 
  \begin{subfigure}{.3\linewidth}
    \centering
        \includegraphics[width=\linewidth]{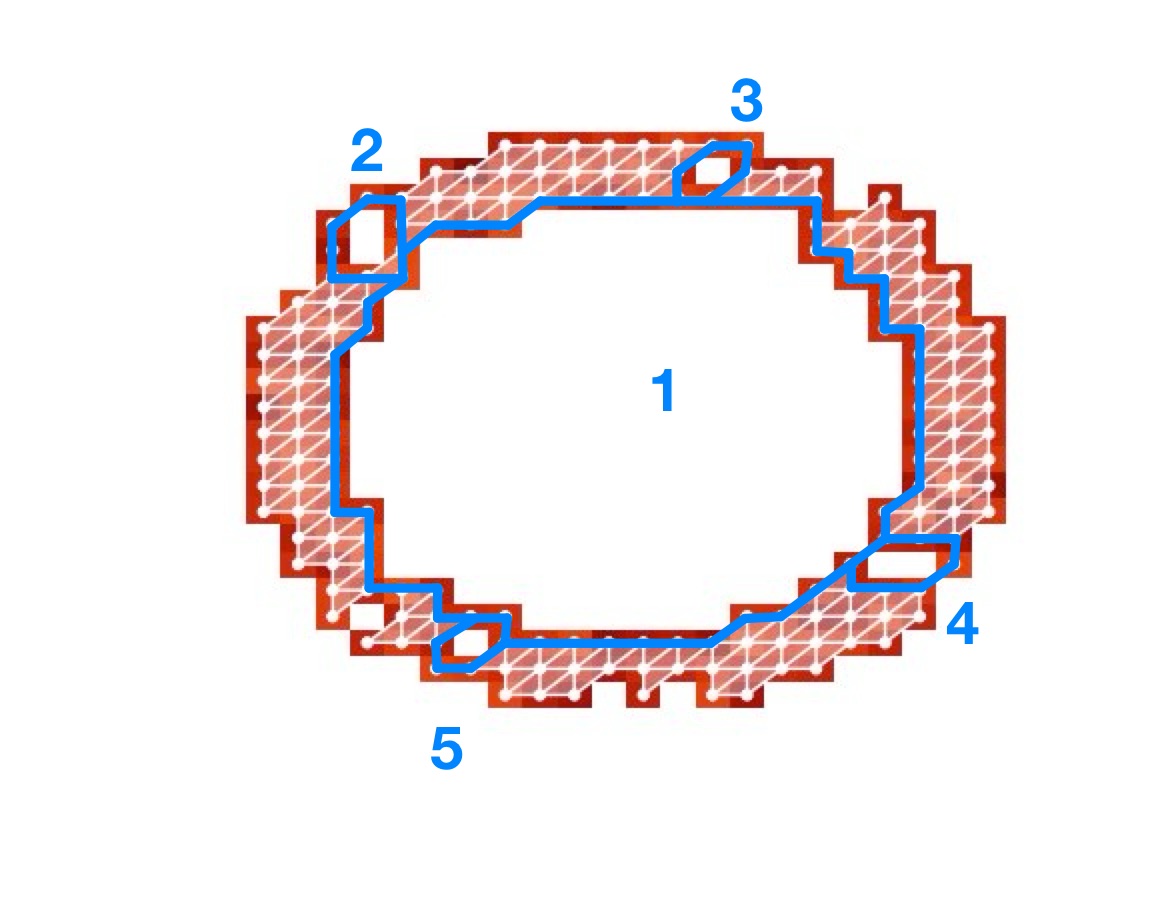}
        \caption{$\delta$=3778 (birth)}\label{subfig:filtration1}
  \end{subfigure}
    \begin{subfigure}{.3\linewidth}
    \centering
        \includegraphics[width=\linewidth]{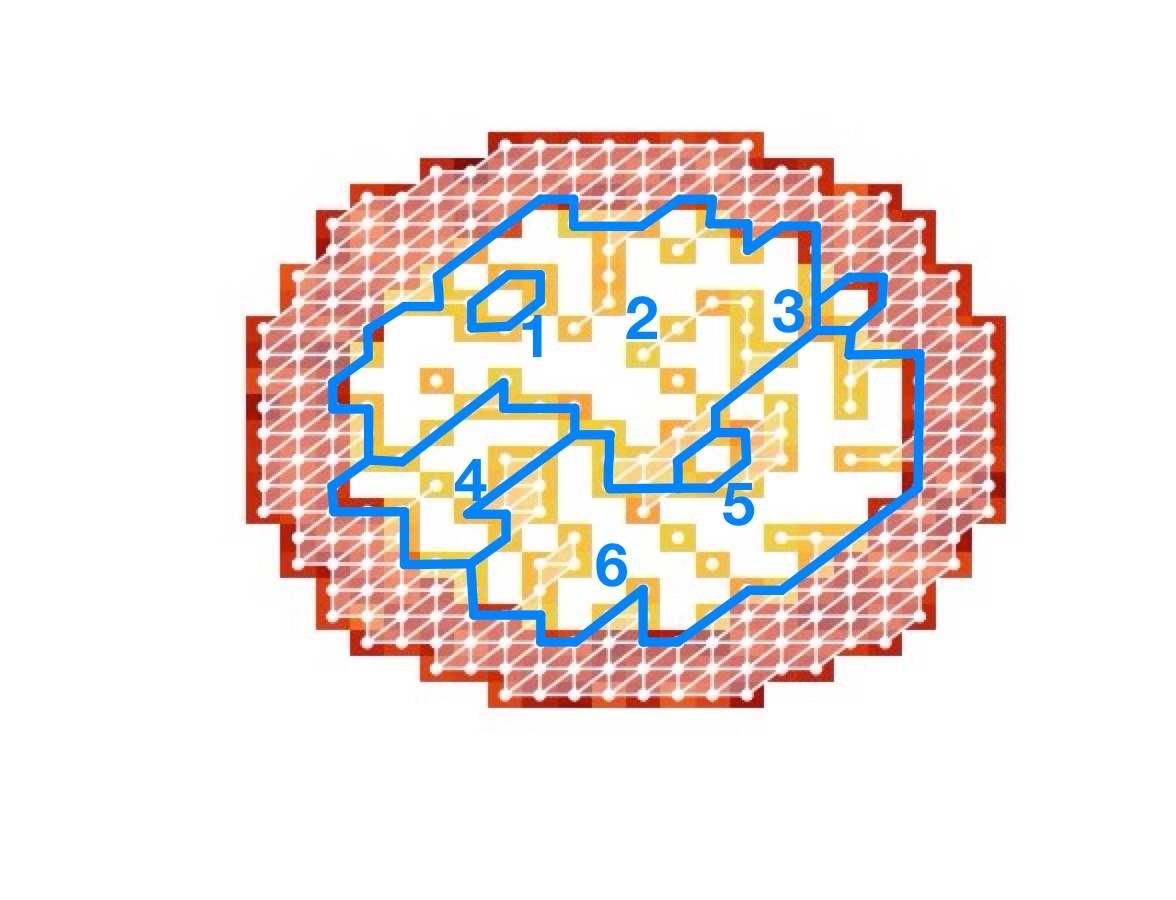}
         \caption{$\delta$=3000}\label{subfig:filtration2}
  \end{subfigure}
  \begin{subfigure}{.3\linewidth}
    \centering
        \includegraphics[width=\linewidth]{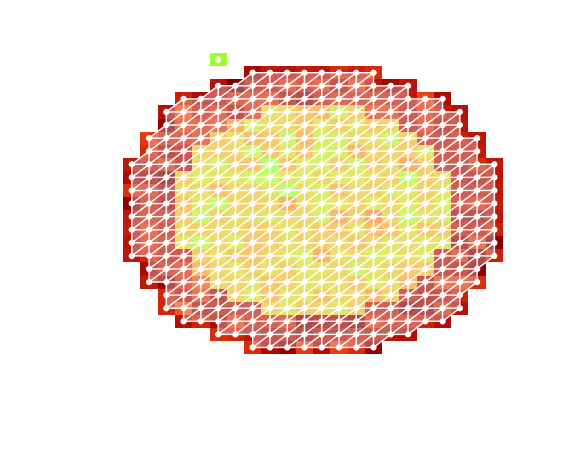}
         \caption{$\delta$=2512 (death)}\label{subfig:filtration3}
  \end{subfigure}
\caption{(A) Partitions of an underlying simulated pattern into background $\mu_0$, loop $\mu_1$, and interior of the loop $\mu_{1*}$. (B) Partitions of the data image (observed) into into background, the loop, and the interior of the loop. Each $Z^k$ denotes pixel intensity value for partition $k=\{0,1,1*\}$ and each $F_k$ is a distribution from which the pixel was drawn. (C) The simplicial complex $\mathcal{K}_{3778}$ built on $(\mathcal{M}^{\sigma})^{[3778,\infty)}$ contains one connected component and five loops; 3778 is the birth time of the true loop (loop number 1). (D) The simplicial complex $\mathcal{K}_{3000}$ built on $(\mathcal{M}^{\sigma})^{[3000,\infty)}$ contains five connected components and six loops. (E) The simplicial complex $\mathcal{K}_{2512}$ built on $(\mathcal{M}^{\sigma})^{[2512,\infty)}$ contains two connected components and no loops, where 2512 is the death time of the large loop born at $\mathcal{K}_{3778}$.}\label{fig:Filtration}
\end{figure}

When the topological space is an image $\mathcal{M}$, the scales can refer to the intensity values of pixels $Z(x,y)$ where the $(x,y)$ coordinates represent the locations of the center of the pixels in the image. Homology groups at different intensities are computed from a triangulation on the upper-level sets of the image, defined as $\mathcal{M}^{-1}(\delta , \infty) = \{(x,y) \in \mathbb{R}^2 \vert Z(x,y) > \delta \}$ where $\delta$ is the threshold for intensity values \cite{15}. This triangulation breaks down the space into simplices—geometric elements on which the computations are carried out. A simplicial complex $\mathcal{K}$ is a set composed of zero-simplices (points), one-simplices (line segments), and two-simplices (triangles), such that (i) any face of a simplex of $\mathcal{K}$ is also a simplex in $\mathcal{K}$, and (ii) the intersection of any two simplices in $\mathcal{K}$ is a face of both simplices or empty. Let $V$ be the set of points ($(x,y)-$coordinates) and $K$ be the set of line segments and triangles which make up $\mathcal{K}$. When a pixel is in $\mathcal{M}^{-1}(\delta , \infty)$, a zero-simplex is placed at the center of that grid cell. A Freudenthal triangulation is then applied to the grid, connecting each zero-simplex to adjacent zero-simplices by one-simplices where adjacency includes horizontal, vertical, and diagonal (bottom-left to top-right direction only) neighbors; see Figure~\ref{fig:Filtration} for an illustration. The pairwise connection of three zero-simplices form a two-simplex \cite{15, 17}.

Figure~\ref{fig:Filtration} shows several examples of simplicial complexes built on upper-level sets of a simulated image along with the correct segmentation of the image and the underlying pattern from which the image was generated (e.g., partitions an image into background and manifold(s), details are discussed in Section~\ref{sec:method}). As the threshold parameter $\delta$ decreases from positive infinity to zero, the space becomes more connected, capturing the homology of each simplicial complex. While $\delta$ varies, a {\em filtration} is formed by a finite sequence of nested sub-complexes $\mathcal{K}_{\delta_1} \subset \mathcal{K}_{\delta_2} \subset \ldots \subset \mathcal{K}_{\delta_l} = \mathcal{K}$ where $\delta_i$ for $i = \{1, \ldots, l\}$ are values of the threshold parameter such that $\delta_i < \delta_j$ if $i < j$. Figures~\ref{subfig:filtration1}-\ref{subfig:filtration3} illustrate different $\mathcal{K}_{\delta}$ on the upper-level sets in a filtration of $\mathcal{M}$. The `birth time' $b$ of a loop, is the value of $\delta$ when it first appears in the filtration (e.g., Figure~\ref{subfig:filtration1}), and its `death time' $d$ is the value at which it merges with another feature (e.g., Figure~\ref{subfig:filtration3}). Persistence, defined as the feature's lifetime, is computed as $b - d$. (Note that in a lower-level set filtration birth occurs before death, so persistence is computed as $d-b$ in order to remain positive.) A common interpretation is that features with longer lifetimes represent topological signal, while those with shorter lifetimes are more likely to be topological noise. \cite{14}.

The evolution of the homology groups of $\mathcal{M}$ over the course of the filtration is graphically represented on a persistence diagram $\mathcal{P}(\mathcal{M})$ defined as follows:
\begin{definition}
    Given a function $\mathcal{M}:\mathbb{R}^2 \longrightarrow \mathbb{R}$ defined on a triangulable subspace of $\mathbb{R}^2$, the {\em $H_p$-persistence diagram} $\mathcal{P}_p(\mathcal{M})$ is the multi-set of points in the extended plane $(\mathbb{R} \cup \{+ \infty \}) \times (\mathbb{R} \cup \{+ \infty \})$, where $p$ is the homology (dimension). Each point $(d, b)$ in the diagram represents a homology group generator that existed in $H_p \left(\mathcal{M}^{-1} [\varepsilon, \infty)\right)$ for $\varepsilon \in$ $(d, b]$.
\end{definition}

\begin{figure}
\centering
      \begin{subfigure}{.4\linewidth}
    \centering
        \includegraphics[width=\linewidth]{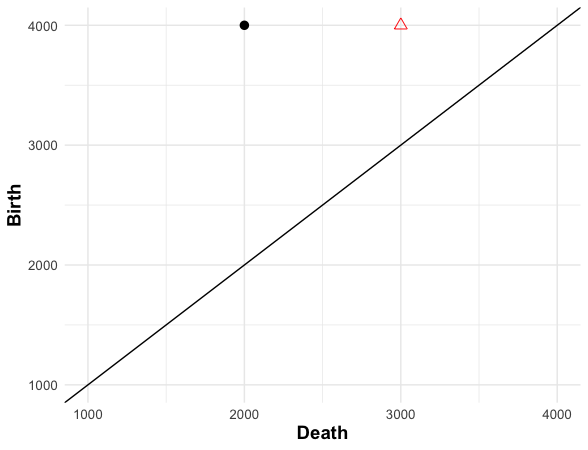}
        \caption{Pattern $\mathcal{P}(\mathcal{M}^{0})$}\label{subfig:PDpattern}
  \end{subfigure}
  \begin{subfigure}{.4\linewidth}
    \centering
        \includegraphics[width=\linewidth]{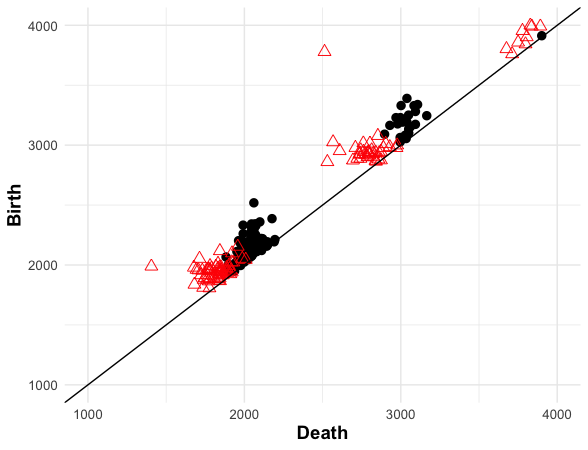}
         \caption{Observed $\mathcal{P}(\mathcal{M}^{\sigma})$}\label{subfig:PDobservered}
  \end{subfigure}
  \begin{subfigure}{.1\linewidth}
    \includegraphics[width=\linewidth]{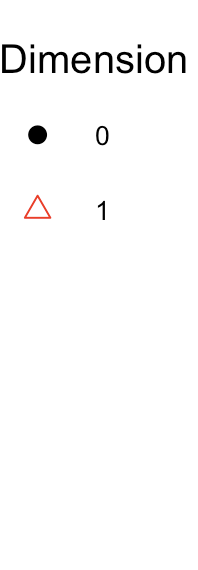}
   \end{subfigure}
\caption{(A) The persistence diagram of the underlying pattern in Figure~\ref{subfig:Pattern} which has only one loop (red triangle) and one connected component (black dots). (B) The persistence diagram of the data in Figure~\ref{subfig:Observered} with loops (red triangles) and connected components (black dots).} \label{fig:PDs}
\end{figure}

Figure~\ref{fig:PDs} displays example persistence diagrams where Figure~\ref{subfig:PDpattern} is the persistence diagram of the underlying pattern from Figure~\ref{subfig:Pattern} and Figure~\ref{subfig:PDobservered} is the persistence diagram of the image from Figure~\ref{subfig:Observered}. Features of each dimension, such as connected components and loops are represented in the diagram by displaying the death and birth times as $(d,b)$ coordinates. Each homology group is represented by a shape and color: connected components are black dots and loops are red triangles. The number of red triangles in each diagram is the number of $H_1$ features (loops) detected in the upper-level set filtration for an image. The more persistent loops are farther from the diagonal line $b=d$.

In the persistence diagram for the data (Figure~\ref{subfig:PDobservered}), the death time of the most persistent loop is $2512$ and the birth time is $3778$, both of these are estimates of the death and birth time of the corresponding loop in the underlying pattern. All the other loops which are closer to the diagonal are small loops which are just due to noise. In the persistence diagram of the underlying pattern (Figure~\ref{subfig:PDpattern}) there is only one loop detected (true loop) with a birth time of $4000$ and a death time of $3000$. 

In the context of our cell biology application, a persistence diagram may be viewed as an estimate of the underlying pattern of $\mathcal{M}$, where a different realization of the image for the same data-generating process generally results in a different persistence diagram. The number of loops, and their corresponding death and birth times, can be viewed as an estimate of the pattern of the ring structure. In the next section, we outline the proposed method for obtaining uncertainty estimates for the death and birth times of the loops found in the data which allows for inference on the true persistence diagram (e.g., Figure~\ref{subfig:PDpattern}) from the observed persistence diagram (e.g., Figure~\ref{subfig:PDobservered}).

\section{Confidence regions for persistence diagram}\label{sec:method}

In this section, we develop a method to assess the uncertainty in the estimated persistence diagram by constructing confidence regions around the death and birth times of the elements in $H_{1}(\mathcal{M^{\sigma}})$, the generators of the one-dimensional homology groups (i.e., loops). These confidence regions should cover the $H_1$ features of the persistence diagram of the noiseless true manifold $\mathcal{M}^0$. However, as is demonstrated in Section~\ref{sec:simulations}, there is considerable bias in the estimated death and birth times of loops using upper-level set filtrations for a raw image, which we refer to as the \emph{traditional TDA} (\textbf{tTDA}) estimates. Bias refers to the systematic deviation of an estimator's expected value from the true value of the parameter being estimated. In this context, if an estimate of a topological feature’s death and birth times in the image is unbiased then it will, on average, match the true death and birth times of the corresponding feature in the underlying pattern. For example, if the goal is to estimate $\mu$ with $\hat b$ then the bias is equal to $\text{Bias}(\hat b) =\mathbb{E}(\hat b)-\mu$ where $\mathbb{E}$ is expected value. 

An approach for reducing the influence of outliers when estimating persistence diagrams for point-cloud data uses upper-level set filtrations on kernel density estimates or regression models of the data, rather than a different type of filtration (e.g., a Vietoris-Rips filtration) on the point-cloud data directly \cite{15, 14}. This technique is used in \cite{14} to construct confidence regions on persistence diagrams for point-cloud data. One important distinction is that a biased estimate is not the same as an outlier. While smoothing can help lessen the impact of outliers, it does not ensure that the death and birth times are unbiased estimates. In fact, smoothing can lead to biased estimates in some settings \cite{dakurah2025maxtda}.

Since the confidence regions are centered around the estimated death and birth times, we need to obtain unbiased estimates of the death and birth times of loops in images. One possible approach, outlined in Section~\ref{sec:comparison}, is to estimate a smoother function of the image and then do an upper-level set filtration extending the inference approach in \cite{14} from point-cloud data to a single image. We refer to this proposed extension as \emph{smooth TDA} (\textbf{sTDA}) which we use as a comparison to our primary proposed approach which we refer to as \emph{partitioned TDA} (\textbf{parTDA}). The \textbf{parTDA} method mitigates the bias apparent in the {\bf tTDA} estimates and also allows for uncertainty quantification without smoothing, and is presented in detail below in Section~\ref{sec:oneH1}.

\subsection{Setup}\label{sec:setup}
Let the image $\mathcal{M}$ be defined by some function $f(x,y)$ discretized onto a 2D grid $\mathcal{G}_{d_1 \times d_2}$, where each $(x,y)$ coordinate represents the grid columns $x=\{1, 2, \ldots, d_1\}$ and  grid rows $y=\{1, 2, \ldots, d_2\}$. The true pattern is the noiseless image $\mathcal{M}^{0} = \{f(x,y) : (x,y) \in \mathcal{G} \}$. However, in practice there is some zero-centered noise $\varepsilon(x,y)$ drawn from distribution $\mathbf{F}(0,\sigma^2(x,y))$ added to the function so that $\mathcal{M}^{\sigma} = \{f(x,y)+\varepsilon(x,y) : (x,y) \in \mathcal{G} \}$ where the $\sigma$ in the exponent indicates there is noise in the image and where $\mathbf{F}$ is some symmetric distribution. Each grid value, or pixel, in $\mathcal{M}^{\sigma}$ has intensity $Z(x,y)$ drawn from: 
\begin{equation}\label{eq:pixel}
    Z(x,y) \sim \mathbf{F}(f(x,y), \sigma^2(x,y)),
\end{equation}
where the mean is defined by $\mathcal{M}^0$ and the error is defined by $\varepsilon$.

In this work, the following assumptions are made regarding the topological features of the noise-free image, $\mathcal{M}^{0}$, which are estimated from the topological features of its noisy counterpart, $\mathcal{M}^{\sigma}$. The proposed method involves partitioning the image in a way that distinguishes the background and $n_p$ other topological structures (e.g., loops and the interior of loops). These assumptions are similar to \cite{14} which assumes uniform sampling on the manifold for point clouds. In general, these assumptions do not apply to smoothed images because neighboring pixels are typically not independent after the smoothing process is applied.

\begin{assumption}\label{as:ContigousRegions} 
The true image $\mathcal{M}^{0}$ can be segmented into contiguous regions with constant functional values: $f(x,y) = \mu_k$ $\forall$ $(x,y)$ within partition $\mathcal{G}_k$. Image $\mathcal{M}^{\sigma}$ can be segmented into $n_p + 1$ (where $n_p$ is the number of partitions corresponding to the topological structures and the plus $1$ is for the background partition) contiguous regions where each region is defined as $\mathcal{M}^{\sigma}_k= \{f(x,y)+\varepsilon (x,y) : (x,y) \in \mathcal{G}_k \}$ for $k= \{ 0, \ldots, n_p \}$ where $\mathcal{G}_k = \{(x,y) \in \mathcal{G} : f(x,y) = \mu_k \}$. 
\end{assumption}

\begin{assumption}\label{as:Homeomorphic1sphere}
If the true image, $\mathcal{M}^{0}$, has at least one feature that is homeomorphic to a one-sphere (loop), let $n_1$ denote the number of one-spheres. Any partition of $\mathcal{M}^{0}$ that is homeomorphic to a one-sphere has pixel intensities fixed at $f(x,y)=\mu_i$ for $i=\{1, \ldots, n_1 \}$ where $2n_1 \leq n_p$ (where $2n_1$ is the number of loops and each loop's interiors), and the partition interior to this one-sphere has pixel intensities fixed at $f(x,y)=\mu_{i*}$. Let $\mu_0$ be designated as the mean of the background noise partition (if it exists). 
\end{assumption}

\begin{assumption}\label{as:UpperLevelSets}
For an upper-level set filtration assume for the majority of $i=\{1, \ldots, n_1 \}$ that $\mu_i \geq \mu_{i*}$ and  $\mu_i \geq \mu_0$. 
\end{assumption}
If all the inequalities from Assumption~\ref{as:UpperLevelSets} are $\geq$, for a given setting, then an upper-level set filtration is sufficient. However, depending on how many $\mu_i \leq \mu_{i*}$, a lower-level set filtration may capture the topological features more effectively. These assumptions enable us to divide an image where the pixels in each region are sampled from the same distribution and exhibit no spatial dependencies.

In Section~\ref{sec:oneH1}, we develop the method to build confidence regions for an image with a single $H_1$ feature (i.e., loop) so that $n_p + 1 = 3$ ($H_1$ feature, the region interior to the $H_1$ feature, and background). A discussion of the generalization of the proposed method to multiple $H_1$ features is presented in  Section~\ref{sec:conc}. An approach for partitioning an image is presented in Section~\ref{sec:segmentation}. Since we were unable to find a method of comparison in the literature, we propose an alternative method in Section~\ref{sec:comparison} that extends the confidence region methodology of \cite{14} from point cloud data to an image. This alternative method ({\bf sTDA}) is used as a benchmark to compare to our proposed method ({\bf parTDA}) in Section~\ref{sec:simulations}.

\subsection{Confidence regions for an image with one \texorpdfstring{$H_1$}{H1} feature}\label{sec:oneH1}

Here we consider the setting with a single loop in $\mathcal{M}^0$. Assumptions~\ref{as:ContigousRegions} and \ref{as:Homeomorphic1sphere} imply that $\mathcal{M}^0$ can be segmented into three contiguous regions where the set of pixel intensities in the background region is defined as $\mathcal{M}_0^0=\{\mu_0: (x,y) \in \mathcal{G}_0 \}$, the set of pixel intensities of the image homeomorphic to a one-sphere is defined as $\mathcal{M}_1^0=\{\mu_1: (x,y) \in \mathcal{G}_1 \}$, and set of pixel intensities that is interior to this one-sphere is defined as $\mathcal{M}_{1*}^0=\{\mu_{1*}: (x,y) \in \mathcal{G}_{1*} \}$. For Section~\ref{sec:oneH1}, we assume the true partitions $\mathcal{G}_0$, $\mathcal{G}_1$, and  $\mathcal{G}_{1*}$ are known. However, in practice, the true partitions are unknown, and segmentation is used to estimate each $\mathcal{G}_k$. Section~\ref{sec:segmentation} proposes an algorithm for reducing the number of misclassified pixels in an estimated segmentation.

Using the known partitions, the data $\mathcal{M}^{\sigma}$ can be separated into three sets of pixels ($\mathcal{M}^{\sigma}_0$, $\mathcal{M}^{\sigma}_1$, $\mathcal{M}^{\sigma}_{1*}$) where the pixels within each set are drawn from distributions  as defined in Assumption~\ref{as:ContigousRegions}: 
\begin{align} \label{eq:paritionsData}
    \mathcal{G}_0 \text{ is the background partition where pixels } Z^0 &\sim \mathbf{F_0}(\mu_{0},\sigma^2_{0}) \nonumber \\
    &\text{ for } Z^0 \in \mathcal{M}^{\sigma}_0 \nonumber \\
    \mathcal{G}_1 \text{ is the part homeomorphic to a one-sphere where pixels } Z^1 &\sim \mathbf{F_{birth}}(\mu_{1},\sigma^2_{1}) \nonumber \\
    & \text{ for } Z^1 \in \mathcal{M}^{\sigma}_1 \nonumber \\
    \mathcal{G}_{1*} \text{ is the part interior to the one-sphere where pixels } Z^{1*} &\sim \mathbf{F_{death}}(\mu_{1*},\sigma^2_{1*}) \nonumber \\
    & \text{ for } Z^{1*} \in \mathcal{M}^{\sigma}_{1*} 
\end{align}

The loop in the true pattern, of which we are trying to estimate the true death and birth times, has a death time of $\mu_{1*}$ and a birth time of $\mu_1$, as shown in Figure~\ref{subfig:Pattern}. We use the sample mean of pixels in each partition $\mathcal{G}_{1*}$ and $\mathcal{G}_{1}$ as unbiased estimates of true the death and birth times of the loop (i.e., the expected values of the sample means are as follows: $\mathbb{E}(\bar Z^{1*}) = \mathbb{E}(\sum_{i=1}^{n_d} Z^{1*}_i)/n_d = \mu_{1*}$ and $\mathbb{E}(\bar Z^{1})  = \mathbb{E}(\sum_{i=1}^{n_b} Z^{1}_i)/n_b = \mu_{1}$) where $n_d$ and $n_b$ are the number of pixels in $\mathcal{M}_{1*}^{\sigma}$ and $\mathcal{M}_{1}^{\sigma}$, respectively. This is possible since all the pixels within the death and birth partitions are independent and identically distributed from each other ($Z^{1*}_i \overset{iid}{\sim} \textbf{F}_\textbf{death} $ and $Z^1_i \overset{iid}{\sim} \textbf{F}_\textbf{birth}$). In order to make the confidence regions, we define the bivariate distribution of the sample means of the pixel intensities associated with the death and birth time estimates, $\bar Z^{1*}$ and $\bar Z^{1}$, respectively, as follows:
\begin{equation} \label{eq:BivariateNormal}
\mathbf{X} = \begin{pmatrix}
\bar Z^{1*} \\
\bar Z^{1} \\
\end{pmatrix} 
\overset{\text{approx}}{\sim}
\left(
\begin{pmatrix}
\mu_{1*} \\
\mu_{1} 
\end{pmatrix} ,
\begin{pmatrix}
\frac{\sigma_{1*}^2}{n_d} & 0 \\
0 & \frac{\sigma_{1}^2}{n_b}
\end{pmatrix}
\right),
\end{equation}
where the first vector on the right indicates the bivariate expected value and the second term is the $2\times2$ covariance matrix.
By the Central Limit Theorem, $\mathbf{X}$ approximately follows a bivariate normal distribution allowing for a confidence region to be created based on: 
$\boldsymbol{(X - \mu)^T \Sigma^{-1} (X - \mu)} \sim \boldsymbol{\chi^2_2} $. The asymptotic confidence region for the death and birth times of $\mathcal{M}^0$ is as follows:
\begin{equation}\label{eq:CI}
    \boldsymbol{\mu}(\theta) = \boldsymbol{X} + \sqrt{\boldsymbol{\chi^2_{2,\alpha}}} \sqrt{\boldsymbol{\hat \Sigma}} \begin{bmatrix} \text{cos}(\theta) \\  \text{sin}(\theta) \end{bmatrix} \text{ for } 0 < \theta < 2\pi,
\end{equation} 
where the variance can be estimated by sample variance of the pixels in each partition, $\mathcal{G}_{1*}$ and $\mathcal{G}_{1}$. The application of the Central Limit Theorem requires Assumptions~\ref{as:ContigousRegions} and \ref{as:Homeomorphic1sphere} stated above; specifically, the pixel intensities within the partitions of the loop and the interior of the loop must each be independent and identically distributed with no spatial correlations.

The segmentation of $\mathcal{M}^{\sigma}_k$ for $k=\{0,1,1*\}$ creates the confidence regions in Equation~\eqref{eq:CI} and the unbiased estimators for $(\mu_{1*},\mu_1): (\bar Z_{1*}, \bar Z_1)$. However, these unbiased estimates are not derived from an upper-level set filtration on $\mathcal{M}^{\sigma}$. We call this approach for generating confidence regions \textbf{parTDA}; next we describe the bias in \textbf{tTDA} methods.
\subsection{Bias in traditional TDA death and birth times}\label{subsubsec:bias}
The level of bias in the {\bf tTDA} birth time is dependent on the proportion of the number of vertices of the simplicial complex that comprise the birth of the loop that are within the set of pixels associated with the corresponding true loop pattern. A similar bias is found with the {\bf tTDA} death time and the relationship of the structure of the simplicial complex and the interior of the true pattern. A more technical explanation is provided next.

\begin{figure}
\centering
  \begin{subfigure}{.45\linewidth}
        \includegraphics[width=\linewidth]{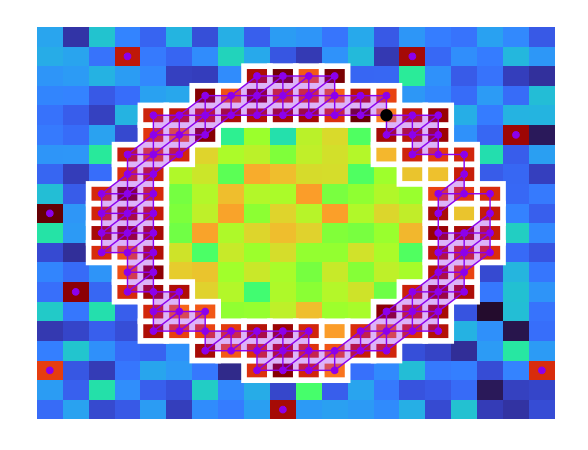}
        \caption{$\mathcal{K}_{b_j}$ on $(\mathcal{M}^{\sigma})^{-1}(b_j,\infty)$}\label{subfig:thinBirth}
  \end{subfigure}
    \begin{subfigure}{.45\linewidth}
        \includegraphics[width=\linewidth]{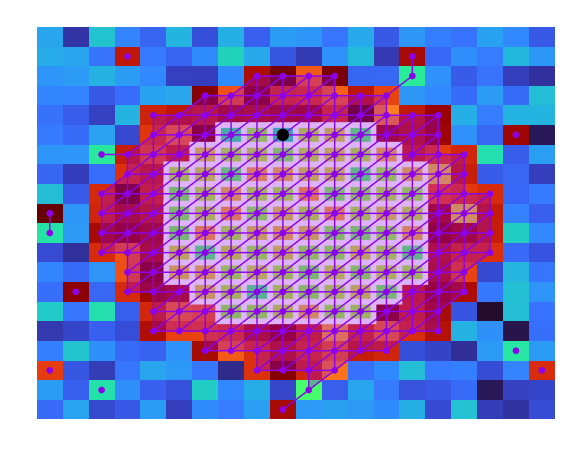}
        \caption{$\mathcal{K}_{d_j}$ on $(\mathcal{M}^{\sigma})^{-1}(d_j,\infty)$}\label{subfig:thinDeath}
  \end{subfigure}
      \begin{subfigure}{.07\linewidth}
        \includegraphics[width=\linewidth]{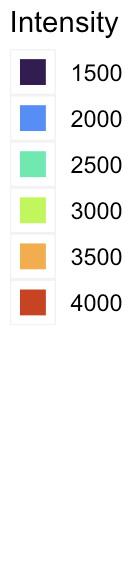}
  \end{subfigure}
\caption{Illustration of simplicial complexes (purple) built on the upper-level sets at the birth time and death time of an image with a loop, with $\mu_1=4000$ and $\mu_{1*}=3000$. (A) The simplicial complex $\mathcal{K}_{b_j} = \mathcal{K}_{3597}$ at the birth of the loop where the black dot is the pixel with intensity value equal to 3597, which is the upper-level set threshold associated with the birth of the loop. The white rectangles indicate the pixels of $\mathcal{M}^{\sigma}_1$. (B) The simplicial complex $\mathcal{K}_{d_j} = \mathcal{K}_{2593}$ at the death of the loop where the black dot is the pixel with intensity value equal to 2593, which is the upper-level set threshold associated with the death of the loop. The white rectangles, which indicate the pixels of $\mathcal{M}^{\sigma}_{1*}$, appear light purple due to the overlaying two-simplices in $\mathcal{K}_{d_j}$.
}\label{fig:LoopConstruction}
\end{figure}

Assumption~\ref{as:UpperLevelSets} states that $\mu_1 \geq \mu_0$ and $\mu_1 \geq  \mu_{1*}$. When applying an upper-level set filtration to $\mathcal{M}^{\sigma}$, a number of loops can be identified along with their associated death and birth times $\{(d_1, b_1), \dots, (d_j, b_j), \ldots (d_{\beta_1}, b_{\beta_1})\}$. Let $\beta_1$ be the total number of loops detected and $(d_j, b_j)$ be the {\bf tTDA} death and birth times for the loop $\mathcal{M}^{\sigma}_1$ (topological signal); all other death and birth times are topological noise and not a part of the true pattern $\mathcal{M}^0$. The birth time, $b_j$, is the smallest $\delta$ value in the filtration when the loop in $\mathcal{M}^{\sigma}_1$ first appears in the simplicial complex $\mathcal{K}_{b_j}=\{V_{b_j},K_{b_j}\}$ where $V_{b_j}$ are the set of $(x,y)-$coordinates in the upper-level set $(\mathcal{M}^{\sigma})^{[b_j,\infty)}$ and $K_{b_j}$ are the higher order simplices connecting the vertices in $V_{b_j}$. The part of the simplicial complex, $\mathcal{K}_{b_j}$, that comprises the birth of the loop is defined as follows:
\begin{equation} \label{eq:paritionsSC_birth}
    \mathcal{K}_{\text{birth}}=\{V_{\text{birth}},K_{\text{birth}}\} \subseteq \mathcal{K}_{b_j} \text{ and } V_{\text{birth}} \subseteq \mathcal{G}_{1}. 
\end{equation}
Similarly, the death time, $d_j$, is the largest $\delta$ value in the filtration when the loop in $\mathcal{M}^{\sigma}_1$ disappears in the simplicial complex $\mathcal{K}_{d_j}=\{V_{d_j},K_{d_j}\}$. The part of the simplicial complex, $\mathcal{K}_{d_j}$, that makes up the interior of the loop is defined as follows:
\begin{equation} \label{eq:paritionsSC_death}
    \mathcal{K}_{\text{death}}=\{V_{\text{death}},K_{\text{death}}\} \subseteq \mathcal{K}_{d_j} \text{ and } V_{\text{death}}=\mathcal{G}_{1*}. 
\end{equation}
Figure~\ref{fig:LoopConstruction} illustrates the difference between $b_j, d_j, \mathcal{K}_{b_j}, \mathcal{K}_{\text{birth}}, \mathcal{K}_{d_j}, \mathcal{K}_{\text{death}}, \mathcal{M}^{\sigma}_1, $ and $ \mathcal{M}^{\sigma}_{1*}$. The white rectangles in each subfigure outline pixels located in $ \mathcal{G}_{1}$ with intensity values in $\mathcal{M}^{\sigma}_1$(Figure~\ref{subfig:thinBirth}) or pixels located in $ \mathcal{G}_{1*}$ with intensity values in $\mathcal{M}^{\sigma}_{1*}$  (Figure~\ref{subfig:thinDeath}), the total purple simplicial complexes are either $\mathcal{K}_{b_j}$ (Figure~\ref{subfig:thinBirth})  or $\mathcal{K}_{d_j}$ (Figure~\ref{subfig:thinDeath}), while the part of the purple simplicial complexes within the white rectangles are either $\mathcal{K}_\text{birth}$ (Figure~\ref{subfig:thinBirth}) or $\mathcal{K}_\text{death}$ (Figure~\ref{subfig:thinDeath}). The black zero-simplex is the location of the pixel which has intensity $b_j$ (Figure~\ref{subfig:thinBirth}) or $d_j$ (Figure~\ref{subfig:thinDeath}). Note that any of the white rectangles beneath the purple simplicial complex appear light purple.

The level of bias in the estimate of $b_j$ using {\bf tTDA} depends on the proportion between the number of elements in the set $\mathcal{K}_{\text{birth}}$ and the number of elements in the set $\mathcal{G}_{1}$, represented by $p_b$. According to Equation~\eqref{eq:paritionsSC_birth}, where $Z(V_{\text{birth}}) \subseteq \mathcal{M}^{\sigma}_1$ and $Z^1 \sim F_{\textbf{birth}}(\mu_1, \sigma^2_1)$, the proportion $p_b$ is defined as follows:
\begin{equation}\label{eq:p_b}
p_b=1-\dfrac{|V_{\text{birth}}|}{|\mathcal{M}^{\sigma}_1| }=1-\dfrac{|V_{\text{birth}}|}{n_b},
\end{equation}
where $| X |$ is the cardinality of the set $X$. 

\begin{figure}
\centering
    \begin{subfigure}{.44\linewidth}
        \includegraphics[width=\linewidth]{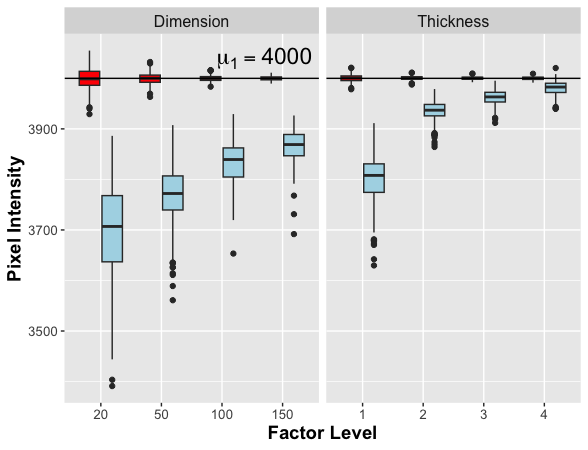}
        \caption{Birth times $(b_j)$}\label{subfig:LoopConstructionBirth}
  \end{subfigure}
  \begin{subfigure}{.44\linewidth}
        \includegraphics[width=\linewidth]{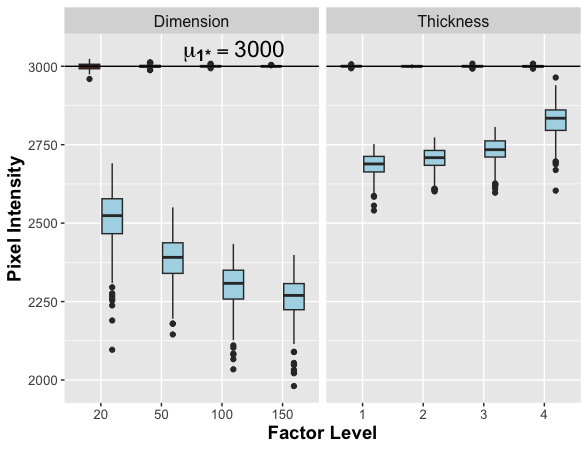}
        \caption{Death Times $(d_j)$}\label{subfig:LoopConstructionDeath}
  \end{subfigure}
    \begin{subfigure}{.1\linewidth}
        \includegraphics[width=\linewidth]{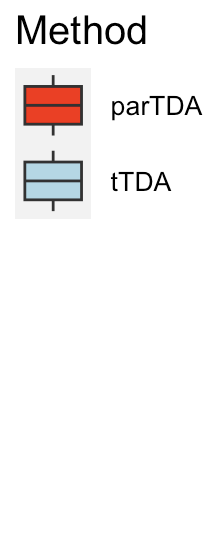}
  \end{subfigure}
\caption{Boxplots illustrating estimated birth (A) and death (B) times of loops using \textbf{parTDA} (red) and \textbf{tTDA} (blue), based on 100 iid images within each factor Level of loop thickness or dimensions. The true death and birth times are indicated by the horizontal solid black lines. The \textbf{tTDA} estimates have a strong negative bias with higher variability, while the proposed \textbf{parTDA} estimates appear to be unbiased with lower variability. } \label{fig:Bias}
\end{figure}

The birth time is an order statistic, representing the minimum intensity value of all the pixels on the simplicial complex of the loop when the loop first appears in the filtration (i.e., $b_j= \min (Z^1(x,y))$ where $(x,y) \in V_{\text{birth}}$). Therefore, it corresponds to some empirical percentile of pixels in the entire loop partition such that $\hat F_{\textbf{birth}}(b_j)=p_b$ where $\hat F_{\textbf{birth}}$ is the empirical distribution function derived from the data. The birth time may correspond to different percentiles of $\hat F_{\textbf{birth}}$ depending on how the image and loop are constructed (e.g., thicker loops may have a birth time closer to the median of the pixel intensities as seen in Figure~\ref{fig:Bias}). The bias in the birth time is defined as:
\begin{equation}
    \text{Bias}(\mu_1,b_j)=\mathbb{E}\{\hat F_{\textbf{birth}}^{-1} (p_b) \}-\mu_1.
\end{equation}
Given the assumption that $F_{\textbf{birth}}$ is a symmetric distribution in Section~\ref{sec:setup}, the {\bf tTDA} birth time is unbiased if $b_j$ is the $50^\text{th}$ percentile of all pixels comprising loop $\mathcal{M}_1^{\sigma}$.

The level of bias of $d_j$ (using {\bf tTDA}) depends on the proportion between the number of elements in the set $\mathcal{K}_{\text{death}}$ and the number of elements in the set $\mathcal{G}_{1*}$, denoted by $p_d$. Based on the Assumptions in Section~\ref{sec:method}, all the pixels which make up the interior of the loop are a part of the simplicial complex at the death of the loop. From Equation~\eqref{eq:paritionsSC_death}, $Z(V_{\text{death}}) \subseteq \mathcal{M}^{\sigma}_{1*}$ and $Z^{1*}\sim F_{\textbf{death}}(\mu_{1*}, \sigma^2_{1*})$, consequently, the proportion $p_d$ is:
\begin{equation} \label{eq:p_d}
 p_d=1-\dfrac{|V_{\text{death}}|}{|\mathcal{M}^{\sigma}_{1*}| } = 1- \dfrac{n_d}{n_d} = 0.
\end{equation}
Then $\hat F_{\textbf{death}}^{-1} (0)= \min(Z^{1*}) = d_j$ where the bias in the estimate is:
\begin{equation} \label{eq:bias}
    \text{Bias}(\mu_{1*},d_j)=\mathbb{E}\{\min(Z^{1*})\}-\mu_{1*}.
\end{equation}
Therefore, the death time is an unbiased estimator of $\mu_{1*}$ when there is only one pixel which makes up $\mathcal{M}^{\sigma}_{1*}$ since, in that case,  $E(\text{min}(Z^{1*}))=E(Z^{1*})=\mu_{1*}$. 

Figure~\ref{fig:Bias} displays results of an exploration of the relationship between bias in {\bf tTDA} estimates of the death and birth times and the construction of the image, compared to the {\bf parTDA} estimates. Differences in the image dimension and the area of the partitions ($\mathcal{G}^{\sigma}_{1*}$ and $\mathcal{G}^{\sigma}_{1}$) change the amount of bias in the {\bf tTDA} estimates of the death and birth times of the loop. Two simulation studies are carried out: (1) considers four different loop thickness levels and (2) considers four different image dimensions levels. Each factor level for both simulations has 100 iid images generated with one loop ($(\mu_{1*}, \mu_1)=(3000,4000)$). 
At each of the loop thickness level, $\{1,2,3,4\}$, the death and birth times of the loop $(d_j,b_j)$ is calculated for each image. Level 1 is for a very thin loop (two pixels thick), level 2 is a medium thin loop (seven pixels thick), level 3 is a medium thick loop (11 pixels thick), and level 4 is for a thick loop (16 pixels thick). Similarly, at each image dimension level, $\{20\times20,50\times50,100\times100,150\times150\}$, the death and birth times of the loop $(d_j,b_j)$ are calculated for each image. These results are shown in boxplots in Figure~\ref{fig:Bias}, where the light blue boxplots are the {\bf tTDA} death and birth times while the red boxplots are the {\bf parTDA} death and birth times.

As seen in Figure~\ref{fig:Bias}, estimates of the birth (Figure~\ref{subfig:LoopConstructionBirth}) and death (Figure~\ref{subfig:LoopConstructionDeath}) times across all different factor levels (dimension and thickness) using {\bf parTDA} appear to be unbiased. Whereas, estimates of the death and birth times using {\bf tTDA} are biased and this bias changes depending on different factor levels. 

In Figure~\ref{fig:Bias}, the dimension of the image serves as a proxy for the pixel sample size of the partitions, with higher dimensions indicating larger sample sizes in both $\mathcal{M}^{\sigma}_{1}$ and $\mathcal{M}^{\sigma}_{1*}$. As image dimension increases, the bias in the birth time estimates using {\bf tTDA} decreases as well as the variance of the birth time estimates. However, for the death time, the bias increases as the image dimensions increase. This result is consistent with the discussion of $p_d$ in Equation~\eqref{eq:p_d}. Loop thickness, which considers the area of $\mathcal{M}^{\sigma}_{1*}$ and $\mathcal{M}^{\sigma}_{1}$, has less bias in both the death and birth time estimates. In general, thicker loops or larger image dimensions (more pixels making up the loop) lead to less biased estimates of the birth time. Thicker loops or smaller image dimensions (fewer pixels making up the interior of the loop) lead to less biased estimates of the death time. In certain situations, the {\bf tTDA} estimates, $(d_j,b_j)$, are unbiased estimators for $(\mu_{1*},\mu_1)$, whereas $(\bar Z_{1*}, \bar Z_1)$ are unbiased regardless of the way the loop or image is constructed. 

\subsubsection{Matching loops between tTDA and parTDA}
The partitions $\mathcal{G}_k$ are used in {\bf parTDA} to estimate the death and birth times of a loop, however, {\bf parTDA} does not detect if a partition forms a loop. Instead, to detect if a partition forms a loop, the unbiased estimates of the death and birth times of a loop, $(\bar Z^{1*}, \bar Z^{1})$, need to be matched to a corresponding loop detected from {\bf tTDA}, $(d_j, b_j)$, for a loop to be detected with {\bf parTDA}. Algorithm~\ref{alg:Seg1H1} is designed to identify which of the loops in the $\mathcal{M}^{\sigma}$, $\{(d_1, b_1), \ldots, (d_j,b_j), \ldots (d_{\beta_1}, b_{\beta_1})\}$, are in the partitions $\mathcal{G}_{1*}$ and $\mathcal{G}_{1}$ by the pixel locations of the death and birth time. The loops which are not matched to the partitions are not considered to be part of the underlying pattern. Once $(d_j,b_j)$ is matched with the partitions $(\mathcal{G}_{1*},\mathcal{G}_1)$ using Algorithm~\ref{alg:Seg1H1}, then the death and birth time estimates of the loop detected with {\bf tTDA} are replaced with $(\bar Z_{1*}, \bar Z_1)$.

\begin{algorithm}
\begin{algorithmic}[1]
\State \textbf{Input: } $df\coloneqq(x, y, Z[x,y])$ of image $\mathcal{M}^{\sigma}$ where $Z[x,y]$ is pixel intensity; partitions $\mathcal{G}_k$ for $k=\{0,1,1*\}$, death and birth times from $\mathcal{P}(M^{\sigma})\coloneqq\{(d_1, b_1), \ldots, (d_{\beta_1}, b_{\beta_1})\}$.
\State \textbf{Output:} $(d_j, b_j)$ matched to $(\mathcal{G}_{1*}, \mathcal{G}_{1})$
\State Define: $df_k =\{(x,y,Z[x,y])  \in \mathcal{G}_k\}$, $k=\{0,1,1*\}$; out$_d=\emptyset$; out$_b=\emptyset$; out$=\emptyset$
    \For{l in 1:$\beta_1$}
        \State{Step 1:} Find $df_k$ where $Z(x,y)=d_l$ \Comment{Identify pixel location of $d_l$ in $\mathcal{G}$}
            \If{$k=1*$} {out$_d\leftarrow$ out$_d\cup l$} \Comment{Only keep index $l$ for $d_l \in \mathcal{G}_{1*}$ }
            \EndIf{}
        \State{Step 2:} Find $df_k$ where $Z(x,y)=b_l$ \Comment{Identify pixel location of $b_l$ in $\mathcal{G}$}
            \If{$k=1$} {out$_b\leftarrow$ out$_b\cup l$} \Comment{Only keep index $l$ for $b_l \in \mathcal{G}_{1}$ }
            \EndIf{}
    \State{Step 3:} Calculate out$\leftarrow$out$_d \cap$ out$_b$
     \If{length($out$)==2} {$(d_l,b_l)=(d_j,b_j)$ \Comment{If $b_l \in \mathcal{G}_{1}$ and $d_l \in \mathcal{G}_{1*}$ loop is matched}}
     \State{{\bf Stop}} 
     \Comment{Match found, stop algorithm}
     \EndIf{}
    \EndFor{}
\State \Return{out}
\end{algorithmic}
\caption{Localizing the death and birth times $(d_j,b_j)$}
\label{alg:Seg1H1}
\end{algorithm}

For each $H_1$ feature $l$ of $\beta_1$ from a {\bf tTDA} persistence diagram of an image $\mathcal{M}^{\sigma}$, Algorithm~\ref{alg:Seg1H1} proceeds with three steps described next.
In Step 1, the algorithm identifies if any of the death times on the persistence diagram match a pixel intensity in the death partition (interior of the loop). The index value of any matched death time is stored in a vector. In Step 2, the algorithm iteratively identifies if any of the birth times on the persistence diagram match a pixel intensity in the birth partition (loop). Then the index value of any matched birth time is stored in a vector. In Step 3, if a death-birth pair $(d_j, b_j)$ from {\bf tTDA} are both equal to a pixel intensity in the death and birth partitions, respectively (i.e., $d_j = Z(x,y) \text{ for } (x,y) \in \mathcal{G}_{1*} \text{ and } b_j = Z(x,y) \text{ for } (x,y) \in \mathcal{G}_{1}$), then that loop is matched to the sample mean intensities of the pixels in those partition $(\bar Z^{1*}, \bar Z^1)$. This process allows us to use the unbiased estimates $(\bar Z^{1*}, \bar Z^1)$ from {\bf parTDA} instead of using biased estimates of the death and birth time from {\bf tTDA}.

\subsection{Confidence regions for multiple \texorpdfstring{$H_1$}{H1} features}

The {\bf parTDA} can be generalized from the setting with only one loop in $\mathcal{M}^0$, which is the setting of our motivating cell image application presented in Section~\ref{sec:CellImages}, to multiple loops in $\mathcal{M}^0$. While the primary emphasis is on $H_1$ features, it is worth noting that the methodology can be readily extended to $p$-spheres for higher-dimensional spaces, such as 3D images, as outlined below.

Assume that there are $n_1$ loops in $\mathcal{M}^0$ resulting in $2n_1 + 1$ partitions and that the functional value of each loop in $f(x,y)$ is $\mu_i$ and the value of the interior of each $H_1$ feature in $f(x,y)$ is $\mu_{i*}$ for $i=\{1, \ldots n_1 \}$. For every loop of $\mathcal{M}^0$, the persistence diagram of the observed image represents each loop as death birth pairs: $(d_{j_1}, b_{j_1}), \ldots (d_{j_{n_1}}, b_{j_{n_1}})$. The steps listed in Algorithm~\ref{alg:Seg1H1} can be extended to connect each $(d_{j_i}, b_{j_i})$ with $(\mathcal{G}_{i*}, \mathcal{G}_{i})$ where the partitions $\mathcal{G}_k$ become $k=\{0,i,i*\}$ for $i=\{1, \ldots, n_1\}$. 

There are three other possible types of birth-death pairs $(d_l,b_l)$ where $l \neq j_i$ for $i=\{1, \ldots n_1 \}$ detected in the image $\mathcal{M}^{\sigma}$ which are not loops in $\mathcal{M}^0$: 

(1) loops which are in the background ($d_0,b_0 \sim \mathbf{F}_{0}(\mu_0, \sigma_0^2)$)
\begin{equation}
    d_0 \notin \mathcal{M}^{\sigma}_{i*} \text{ and } b_0 \notin \mathcal{M}^{\sigma}_{i} \forall i \neq 0 \implies \text{using Algorithm \ref{alg:Seg1H1} } (d_0,b_0)\neq(d_{j_i},b_{j_i})
\end{equation}

(2) loops which are only in $\mathcal{M}_{i}$ or only in $\mathcal{M}_{i*}$ ($d_i,b_i \sim \mathbf{F}_{\textbf{birth}}(\mu_i, \sigma_i^2)$ or $d_{i*},b_{i*} \sim \mathbf{F}_{\textbf{death}}(\mu_{i*}, \sigma_{i*}^2)$)
\begin{align}
    d_i,b_i \in \mathcal{M}^{\sigma}_i  &\implies  \text{using Algorithm~\ref{alg:Seg1H1} } (d_i,b_i)\neq(d_{j_i},b_{j_i}) \\
    d_{i*},b_{i*} \in \mathcal{M}^{\sigma}_{i*} &\implies  \text{using Algorithm~\ref{alg:Seg1H1} } (d_{i*},b_{i*})\neq(d_{j_i},b_{j_i})
\end{align}

(3) loops that traverse the background and $\mathcal{M}_{i}^{\sigma}$ ($b_i \sim \mathbf{F}_{\textbf{birth}}(\mu_i,\sigma^2_i)$ and $d_0 \sim \mathbf{F}_{0}(\mu_0,\sigma^2_0)$)
\begin{equation}
    d_0 \notin \mathcal{M}^{\sigma}_{i*} \implies \text{using Algorithm~\ref{alg:Seg1H1} } (d_0,b_i)\neq(d_{j_i},b_{j_i})
\end{equation}

Since all the loops detected in the segmentation $\mathcal{M}^{\sigma}_i$ are connected to the correct $(d_{j_i}, b_{j_i})$, the only time a problem would arise is when $d_{j_i}=d_{j_k}$ and $b_{j_i}=b_{j_k}$ for $i \neq k$ where $i,k \in \{1, \ldots, n_1\}$. In other words, if the loop $\mathcal{M}^{\sigma}_i$ and the loop in $\mathcal{M}^{\sigma}_k$ have the exact same death and birth times, the algorithm would not be able to match $(d_{j_i}, b_{j_i})$ and $(d_{j_k}, b_{j_k})$ with $(\mathcal{G}_{i*}, \mathcal{G}_{i})$ and  $(\mathcal{G}_{k*}, \mathcal{G}_{k})$, respectively. However, this situation would happen with zero probability since all $Z^{i} \sim \mathbf{F}_{\textbf{birth}}(\mu_i, \sigma_i^2)$, $Z^{k} \sim \mathbf{F}_{\textbf{birth}}(\mu_{k}, \sigma_{k}^2)$, $Z^{i*} \sim \mathbf{F}_{\textbf{death}}(\mu_{i*}, \sigma_{i*}^2)$, and $Z^{k*} \sim \mathbf{F}_{\textbf{death}}(\mu_{k*}, \sigma_{k*}^2)$ are continuous distributions. If the intensity values in the image are discrete rather than continuous (e.g., the integers), one way to address this issue is to smooth the image. Smoothing transforms intensity into continuous values which helps in identifying the location of loops in the image through their unique death and birth times.

\subsection{Segmentation of the image}\label{sec:segmentation}

In the preceding two subsections the partitions $\mathcal{G}_k$ for $k = \{0, \ldots, n_p \}$ are assumed to be known; whereas in this section, the segmentation is unknown and is estimated with $\mathcal{\hat G}_k$ for $k=\{0, \ldots, \hat n_p\}$. If the segmentation is incorrect, the {\bf parTDA} estimated death and birth times in Equation~\eqref{eq:BivariateNormal} and the corresponding confidence regions in Equation~\eqref{eq:CI} may not be accurate. Here, we propose a method to reduce the misclassification of pixels in partitions when one or more of the $\mathcal{\hat G}_k$'s may have some incorrect pixels assigned to it. 

Recall from Equation~\eqref{eq:paritionsData} that if $\mathcal{G}_k$ is known $\forall k \in \{0, \ldots, n_p\} $ then interior pixel intensities $Z^{i*} \sim \textbf{F}_{\textbf{death}} (\mu_{i*},\sigma_{i*}^2)$ for every $Z^{i*} \in \mathcal{M}^{\sigma}_{i*}$ and pattern pixel intensities $Z^i \sim \textbf{F}_{\textbf{birth}} (\mu_i,\sigma_i^2)$ for every $Z^{i} \in \mathcal{M}^{\sigma}_{i}$, where $i \in \{1, \ldots, n_1\}$, with the number of pixels in the sets defined as $|\mathcal{M}^{\sigma}_{i*}|=n_d^i$ and $|\mathcal{M}^{\sigma}_{i}|=n_b^i $. 

When $\mathcal{G}_k$ is unknown, $\mathcal{\hat M}^{\sigma}_{i}$ and $\mathcal{\hat M}^{\sigma}_{i*}$ are estimated using some segmentation procedure. Any segmentation procedure may be used to estimate the partitions, as long as the resulting partitions are contiguous regions. In this paper, we apply edge detection methods to segment the image by identifying edges, which are located at the maxima of the gradient strength obtained from a Laplacian of the Gaussian-smoothed image \cite{18, 19}. For certain parameter values, the edge contours are closed, creating contiguous regions, and the standard deviation of the filter changes how many regions are detected. Let $\hat e$ be the edge set which segments the image $\mathcal{M}^{\sigma}$ into partitions $\mathcal{\hat G}_k$. 

Assume that some part of the segmentation of a loop or its interior is incorrect so that $\mathcal{\hat G}_k \neq \mathcal{G}_k$ for $k =\{i,i*\}$ for some $i$. Then there are $m_d$ pixel intensities, denoted by  $\tilde Z^{i*}$, in the set $\mathcal{M}_{i*}$ which are misclassified into $\mathcal{\hat M}_{i}$ (i.e., these are the pixels that should be a part of the interior, but were assigned to the loop). Similarly, there are $m_b$ pixel intensities, denoted by  $\tilde Z^{i}$, in the set $\mathcal{M}_{i}$ which are misclassified into $\mathcal{\hat M}_{i*}$ (i.e., these are the pixels that should be a part of the loop, but were assigned to the interior). There are then $n_d-m_d$ pixel intensities, denoted by $\tilde{\tilde{Z}}^{i*}$, in the set $\mathcal{M}_{i*}$ which are correctly classified into $\mathcal{\hat M}_{i*}$ and there are $n_b-m_b$ pixel intensities, denoted by $\tilde{\tilde{Z}}^{i}$, in the set $\mathcal{M}_{i}$ which are correctly classified into $\mathcal{\hat M}_{i}$.

The set of pixel intensity values which comprise the true interior of the loop ($\mathcal{M}_{i*}$) and the set of pixel intensity values which comprise the true loop ($\mathcal{M}_{i}$) can be decomposed as follows: 
\begin{align}
    \mathcal{M}_{i*} = \tilde{\tilde{Z}}^{i*} \cup \tilde Z^{i*} \text{ and } \mathcal{M}_{i} = \tilde{\tilde{Z}}^{i} \cup \tilde Z^{i}.
\end{align}

Let $\mathcal{\hat M}_{i*}$ denote all the intensity values of pixels which are classified as interior pixels of the loop $\mathcal{\hat G}_{i*}$ (i.e., $\mathcal{\hat M}_{i*}= \tilde{\tilde{Z}}^{i*} \cup \tilde Z^{i}$) and $\mathcal{\hat M}_{i}$ denote all the intensity values of pixels which are classified as loop pixels $\mathcal{\hat G}_{i*}$ (i.e., $\mathcal{\hat M}_{i}=\tilde{\tilde{Z}}^{i}\cup\tilde Z^{i*}$). 
Therefore $n_b-m_b+m_d$ pixels are in the birth time partition $\mathcal{\hat G}_{i}$ and $n_d-m_d+m_b$ pixels are in the death time partition $\mathcal{\hat G}_{i*}$.

The expected value of the (biased) estimators of the death and birth time using the incorrect partitions of the loop are:
\begin{align}
    \mathbb{E}(\bar{\hat Z}^{i}) = \frac{(n_b-m_b) \mu_{i} + m_d \mu_{i*}}{n_b-m_b+m_d} \text{ and } \mathbb{E}(\bar{\hat Z}^{i*}) = \frac{(n_d-m_d) \mu_{i*} + m_b \mu_{i}}{n_d-m_d+m_b},
\end{align}
where $\bar{\hat Z}^{i}$ and $\bar{\hat Z}^{i*}$ are the sample means of the sets of pixels $\mathcal{\hat M}_{i}$ and $\mathcal{\hat M}_{i*}$, respectively.

By Assumption~\ref{as:UpperLevelSets}, $\mu_{i*} \leq \mu_i$ and assuming that the segmentation $\mathcal{\hat G}_i$ and $\mathcal{\hat G}_{i*}$ are close to the true $\mathcal{G}_i$ and $\mathcal{G}_{i*}$ (i.e., only a few pixels are misclassified), then $m_b < n_d$ and $m_d < n_b$ and any $\tilde Z^i \in \mathcal{\hat M}_{i*}$ and $\tilde Z^{i*} \in \mathcal{\hat M}_{i}$ are neighbors of the edge set $\hat e$ (i.e., $\tilde Z^i, \tilde Z^{i*} \in n_c(\hat e)$ where $c$ is the unit distance between two pixels.

Let $q^i_1,q^{i*}_1$ be the first quantiles and  $q^i_3,q^{i*}_3$ be the third quantiles of  $\textbf{F}_{\textbf{birth}}$, $\textbf{F}_{\textbf{death}},$ respectively. Assume that the noise distribution $\varepsilon(x,y) \sim \mathbf{F}(0, \sigma^2(x,y))$ is symmetric. An assumption of Algorithm~\ref{alg:MisclassifiedPixels} is that the distribution of the interior pixel intensities and the pattern pixel intensities are well-separated, as described in the following.
\begin{assumption}\label{as:MisclassifiedAlg}
   Assume that $(o_i-T(\mu_i)) < (o_i-T(\mu_{i*}))$ and $(o_{i*}-T(\mu_{i*})) < (o_{i*}-T(\mu_{i}))$ where $o_i$ is an outlier in the distribution $\textbf{F}_{\textbf{birth}}$ and $o_{i*}$ is an outlier in the distribution $\textbf{F}_{\textbf{death}}$. $T(\mu_{i*})$ and $T(\mu_{i})$ are the truncated means of  $\textbf{F}_{\textbf{death}}$ and $\textbf{F}_{\textbf{birth}}$ with upper bound $q^{i*}_3+1.5(q^{i*}_3-q^{i*}_1)$ and lower bound $q^{i}_3+1.5(q^{i}_3-q^{i}_1)$, respectively.
\end{assumption}

Under Assumption~\ref{as:MisclassifiedAlg} the pixel intensity distributions for the loop and its interior are sufficiently well-separated—specifically, the minimum values in the birth region are closer to the mean of the truncated birth distribution (outliers are removed) than to the mean of the truncated death distribution—Algorithm~\ref{alg:MisclassifiedPixels} identifies and separates the $m_d$ and $m_b$ misclassified pixels, $\tilde Z^{i*}$ and $\tilde Z^{i}$, into the edge set $\hat e$ and keeps the outliers, $\tilde{\tilde{Z}}^{i*} \sim \mathcal{M}^{\sigma}_{i*}$ and $\tilde{\tilde{Z}}^{i} \sim \mathcal{M}^{\sigma}_{i}$ in the correct segments $\mathcal{\hat M}^{\sigma}_{i*}$ and $\mathcal{\hat M}^{\sigma}_{i}$ respectively.

\begin{remark}\label{remark:MisclassifiedAlg}
    Assumption~\ref{as:MisclassifiedAlg} is a strong assumption but is not necessarily required for the algorithm to work well. Recovering a less stringent bound is a topic of future investigation.
\end{remark}

\begin{algorithm}
\begin{algorithmic}
\State \textbf{Input: } edge set $\hat{e}$; image $\mathcal{M}^{\sigma}$; partitions $\mathcal{\hat G}_1$ and $\mathcal{\hat G}_{1*}$; $c$=pixel side length
\State \textbf{Output:} new edge set $\hat{e}^{\text{new}}$
\State Define: $\mathcal{\hat M}^{\sigma}_i=\{Z^i(x,y)_l : (x,y)_l \in \mathcal{\hat G}_i \}$, $L_i=| \mathcal{\hat M}^{\sigma}_i|,$ $P(Z^i(x,y) \leq q^i_{1})=0.25$, $P(Z^i(x,y) \leq q^i_{3})=0.75$ for $i=\{1,1*\}$; outlier$_i=\emptyset$; outlier.idx$_i=\emptyset$; dist()=Euclidean distance; $e_1=\emptyset$
    \For{i in $\{1,1*\}$}
        \For{l in $1:L_i$} \Comment{Check if $Z^i(x,y)_l$ is an outlier and neighbors an edge in $\mathcal{\hat G}_i$}
        \If{$\left( (Z^i(x,y)_l > q^i_{3}+1.5(q^i_{3}-q^i_{1})) \mid (Z^i(x,y)_l < q^i_{1}-1.5(q^i_{3}-q^i_{1}))\right)$ \& \\  $\left(\exists (\tilde x, \tilde y) \in \hat e~s.t.~\text{dist}((x,y)_l,(\tilde x, \tilde y)) \leq \sqrt{2}c  \right)$}{$\text{ outlier}_i\leftarrow \text{outlier}_i \cup Z^i(x,y)_l$, outlier.idx$_i \leftarrow \text{outlier.idx}_i \cup l$}
        \EndIf{}
        \EndFor{}
    \EndFor{}
    \State{Calculate $\hat \mu_1=\mathcal{\hat M}^{\sigma}_1 \setminus$outlier$_1$ and $\hat \mu_{1*}=\mathcal{\hat M}^{\sigma}_{1*} \setminus$outlier$_{1*}$}
    \Comment{Calculate means without outliers} 
    \For{i in $\{1,1*\}$}
    \For{l in outlier.px$_i$} 
            \If{$\vert Z^i(x,y)_l-\hat \mu_i \vert \geq \vert Z^i(x,y)_l-\hat \mu_{i^c} \vert $} \Comment{$i^c$ is the complement in $\{1,1*\}$ for $i$}
            {$e_1\leftarrow e_1 \cup (x,y)_l$} \Comment{only add $(x,y)_l$ to new edge set $e_1$ if $Z^i(x,y)_l$ is closer to $\hat \mu_{i^c}$}
        \EndIf{}
        \EndFor{}
    \EndFor{}
    \State{$\hat e^{\text{new}} = \hat e \cup e_1$}
\State \Return{$\hat e^{\text{new}}$}
\end{algorithmic}
\caption{Remove Misclassified Pixels from Partition ($\mathcal{G}_1$,$\mathcal{G}_{1*}$)}
\label{alg:MisclassifiedPixels}
\end{algorithm}

In the first For-loop, this algorithm iteratively identifies which pixels are: (1) outliers (Assumption~\ref{as:MisclassifiedAlg}) and (2) next to a pixel in the edge set, for the death partition followed by the birth partition, separately. These pixels are stored in a vector and labeled as outliers in the segmentation scheme. The pixels identified as outliers are removed from the death and birth partitions and then a new means are calculated from the remaining pixels. 

In the second For-loop, the algorithm iteratively checks if the outliers from the original partitions are closer to the new mean of the birth partition or the new mean of the death partition. If any of the outliers from the original death partition are closer to the new mean of the birth partition compared to the new mean of the death partition, they are assigned to the new edge set. If any of the outliers from the original death partition are closer to the new mean of the death partition, they are once again allocated to the death partition. An analogous procedure happens to the outliers in the birth partition. The new edge set is the original edge set with any outlying pixels meeting the above criteria.

\begin{figure}
\centering
  \begin{subfigure}{.46\linewidth}
  \centering
  \includegraphics[width=\linewidth]{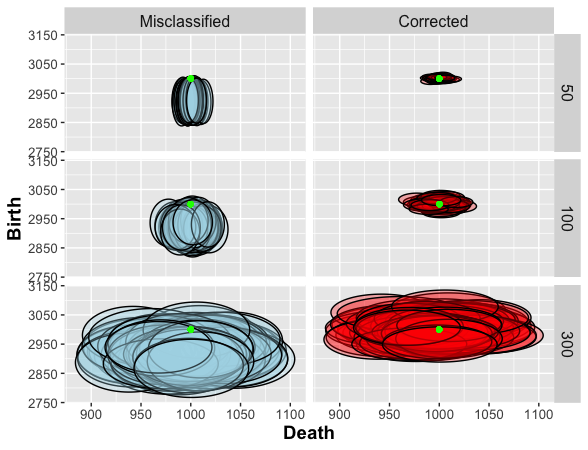}
        \caption{Confidence regions} \label{subfig:MisclassCI}
  \end{subfigure}
  \begin{subfigure}{.47\linewidth}
  \centering
  \includegraphics[width=\linewidth]{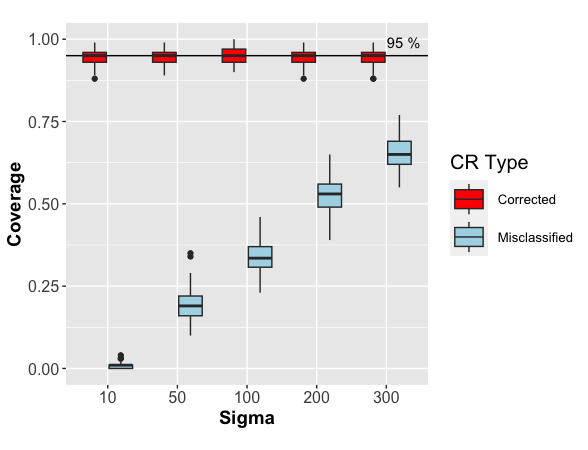}
         \caption{Coverage} \label{subfig:MisclassBoxplots}
   \end{subfigure}
 \caption{Confidence regions and coverage before (misclassified) and after (corrected) Algorithm~\ref{alg:MisclassifiedPixels} has been applied. The misclassified segmentation $\hat e$ has six pixels incorrectly classified. (A) Confidence regions for 100 images at noise level $\sigma=\{50, 100, 300 \}$ are shown using $\hat e$ (misclassified) and $\hat e_{\text{new}}$ (corrected). The green dots indicate the true death and birth time location. (B) The coverage of the 95\% confidence regions for $\sigma=\{10, 50, 100, 200, 300 \}$ for misclassified (red) and corrected (blue) segmentations, using 100 iid images.} \label{fig:MisclassificationAlg}
\end{figure}

As an illustration of the performance of Algorithm~\ref{alg:MisclassifiedPixels}, the following experiment was carried out and results are displayed in Figure~\ref{fig:MisclassificationAlg}. For three different noise settings $(\sigma=\{50,100,300\})$, 100 iid images with one loop, similar to Figure~\ref{subfig:Observered} with $(\mu_{1*},\mu_1)=(1000,3000)$, are generated and segmented incorrectly with the same edge set $\hat e$. In this example, six pixels are misclassified in the loop (i.e., $\tilde Z^{1*} \in \mathcal{\hat G}_{1}$) with the edge set $\hat e$. The  95\% confidence regions using {\bf parTDA} are calculated using both this misclassified partition $\hat e$ and the corrected partion $\hat e_{\text{new}}$ generated from Algorithm~\ref{alg:MisclassifiedPixels}. Lower noise levels have more biased coverage of the resulting confidence regions compared to the higher noise levels.

Figure~\ref{subfig:MisclassCI} shows all 100 estimated 95\% confidence regions built using $\hat e$ (red) and $\hat e_{\text{new}}$ (blue) for the different $\sigma$ values. The green dot is the true $(\mu_{1*},\mu_1)=(1000,3000)$ which the regions should cover $95\%$ of the time, on average. The confidence regions for the misclassified setting are underestimating $\mu_1$ since some $Z^{1*}$ pixel intensities, which are lower than those of $Z^1$, are included in the ${\bar {\hat Z}^1}$ resulting in an estimate that is biased low. After Algorithm~\ref{alg:MisclassifiedPixels} is applied, the bias in the confidence regions appear to be corrected in terms of the birth time.

In Figure~\ref{subfig:MisclassBoxplots}, the coverage is calculated based on 100 iid images at each noise level $(\sigma=\{10,50,100,200,300\})$. The misclassified boxplots (blue) show the coverage of the confidence regions built from $\hat e$, and the corrected boxplots (red) show the coverage for confidence regions calculated with the $\hat e^{new}$ after running Algorithm~\ref{alg:MisclassifiedPixels}. As illustrated in both plots, the algorithm significantly improves the coverage of the confidence regions. Correct segmentation is crucial for {\bf parTDA}, and this analysis emphasizes the importance of checking the segmentation.

\subsection{Alternative method}\label{sec:comparison}

We extend one of the methods from \cite{14} from point-cloud data to handle an image as a way to establish a benchmark in the absence of a direct basis for comparison with \textbf{parTDA}. This approach, which we refer to as \textbf{sTDA}, is used as a comparison to our new method, \textbf{parTDA}. In \textbf{sTDA}, a distribution of distances between the persistence diagrams of the smoothed data, $\mathcal{P}(\mathcal{\tilde M}^{\sigma})$, and the persistence diagram of the true pattern, $\mathcal{P}(\mathcal{\tilde M}^0)$, is used to determine confidence regions on a persistence diagram. 
In particular, persistence diagram stability results \cite{20} are used to bound the (bottleneck) distance between the persistence diagrams by the $L_{\infty}$ distance between kernel density estimates (KDEs) of the point-cloud data and the true pattern. Asymptotic confidence regions are then built from the distribution of $L_{\infty}$ distances between $\mathcal{\tilde M}^{\sigma}$ and $\mathcal{\tilde M}^{0}$, which can be estimated using a bootstrap procedure. 
 
This procedure is briefly outlined below and then followed by the proposed adjustments for image data. See Section~3.4 of \cite{14} for more details.

In the context of \cite{14}, let $\mathcal{M}^{\sigma}$ be point-cloud data. One of their proposed methods for persistence diagram confidence regions considers a KDE of $\mathcal{M}^{\sigma}$, $\mathcal{\tilde M}^{\sigma}$, to estimate the true death and birth time, $(\tilde \mu_{i*}, \tilde \mu_i)$, of the (true) smoothed manifold, $\mathcal{\tilde M}^0$. 
They define an asymptotic $(1-\alpha)100$\% confidence region, adapted to our notation which omits the dependency on bandwidth and sample size; see Theorem 12 of \cite{14} for the precise statements:
\begin{equation}
    \mathbb{P} \left(\mathcal{W}_{\infty} ( \mathcal{P}(\mathcal{\tilde M}^{\sigma}), \mathcal{P}(\mathcal{\tilde M}^0)) > c_n \right) \leq \mathbb{P} \left( \vert \vert \mathcal{\tilde M}^{\sigma} - \mathcal{\tilde M}^0 \vert \vert_{\infty} > c_n \right) \leq \alpha + O \left( n^{-1/2}\right)
\end{equation}
where $c_n$ defines the confidence region based on the data, and the first inequality follows from the stability result of \cite{20}. The bottleneck distance, $\mathcal{W}_{\infty}$ is defined as
\begin{equation}
    W_{\infty}(\mathcal{P}(\mathcal{\tilde M}^{\sigma}), \mathcal{P}(\mathcal{\tilde M}^{0})) = \inf_{\eta: \mathcal{P}(\mathcal{\tilde M}^{\sigma}) \longrightarrow \mathcal{P}(\mathcal{\tilde M}^{0})} \sup_{(d,b)  \in \mathcal{P}(\mathcal{\tilde M})} \vert \vert (d,b) - \eta(d,b) \vert \vert_{\infty},
\end{equation}
where $\eta$ is a bijection of the features of the diagrams, including the diagonal $b=d$ line \cite{20, 14}.
Since $\mathcal{\tilde M}^0$ is unknown and there is only one realization of the data $\mathcal{\tilde M}^{\sigma}$, a bootstrap approach is used. In particular, the estimate of $c_n$ is the $(1-\alpha)$-quantile of the distribution of the $L_{\infty}$ distances between the smoothed data, $\mathcal{\tilde M}^{\sigma}$, and smoothed bootstrap realizations of the point-cloud data. 

To implement this approach for images two modifications are made. First, instead of a KDE on point clouds, we use local polynomial smoothing to change the raw image $\mathcal{M}^\sigma$ into a smoothed image $\mathcal{\tilde M}^\sigma$. In Section~\ref{sec:simulations}, we use degree two polynomials and an adaptive bandwidth of 0.3 as parameter inputs for local polynomial smoothing. These input values resulted in only one loop detected by an upper-level set filtration for the smoothed pattern, $\mathcal{\tilde M}^0$, analogous to the original image, $\mathcal{M}^0$. 

Note that \textbf{sTDA} builds confidence regions to cover $(\tilde \mu_{i*}, \tilde \mu_{i})$ (i.e., death and birth times of loops in $\mathcal{\tilde M}^0$) whereas \textbf{parTDA} builds confidence regions to cover $(\mu_{i*}, \mu_{i})$ (i.e., death and birth times of loops in $\mathcal{M}^0$). So, while the methods are not directly comparable; we can still compare the coverage percentage of the confidence regions for each method. The second modification is creating a new method to bootstrap an image as opposed to a point cloud. The traditional bootstrap method assumes that each observation is iid which is not a suitable assumption for an image which often have spatial correlation. We segment the image into different strata and use the stratified bootstrap to resample the full image. Within each stratum the pixels can be viewed as being drawn from the same distribution, so pixel intensities within each stratum can be bootstrapped. In our simulation study, the number of strata and the segmentation is assumed to be correct for the {\bf sTDA} benchmark.

\section{Simulation study}\label{sec:simulations}

In this section, we empirically evaluate the accuracy, precision, and computational efficiency of the proposed confidence regions. Accuracy is assessed by considering bias in the estimates, empirical coverage percentage, and the identification of the number of loops in the underlying pattern, while precision is evaluated by analyzing the area of the confidence regions. A summary of all of these numerical results are displayed in Table~\ref{tab:simulationResults}. We also report the relative computational efficiency of {\bf parTDA} and {\bf sTDA} in generating confidence regions and estimating the death and birth times of a loop in an image (Table~\ref{tab:simulationTime}). And finally, we evaluate the performance of {\bf parTDA} in the multiple loop setting.

For the simulations, each image has one loop and follows the assumptions from Section~\ref{sec:oneH1}. The death and birth  times of the true pattern, $\mathcal{M}^0$, are set to $(\mu_{1*}, \mu_1)=(1000, 3000)$, which are similar intensities to those of our cell wound example (see Section~\ref{sec:CellImages}). To assess the robustness of the proposed confidence regions to noise, four different noise levels are used to generate an image $\mathcal{M}^{\sigma}$ for $\sigma=\{50, 150, 250, 350 \}$, homoscedastic Gaussian noise is used in this section. For each $\sigma$, $l$ images are generated, denoted $\mathcal{M}^{\sigma}_l$ where $l=\{1, \ldots, 100\}$, and an upper-level set filtration is used to get the death and birth times for each image (i.e., the \textbf{tTDA} estimates). To test the alternative method (\textbf{sTDA}) each image is further smoothed using local polynomial smoothing, denoted $\mathcal{\tilde M}^{\sigma}_l$. Then both \textbf{sTDA} and \textbf{parTDA} are used to get confidence regions for the underlying pattern in $\mathcal{\tilde M}^0$ and $\mathcal{M}^0$, respectively. 

\begin{figure}
\centering
  \begin{subfigure}{.48\linewidth}
    \centering
        \includegraphics[width=\linewidth]{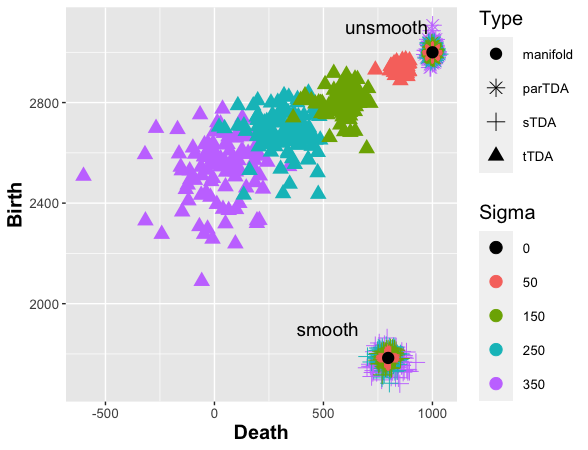}
        \caption{(Death, Birth) of loop} \label{subfig:estimates}
  \end{subfigure}
    \begin{subfigure}{.48\linewidth}
    \centering
        \includegraphics[width=\linewidth]{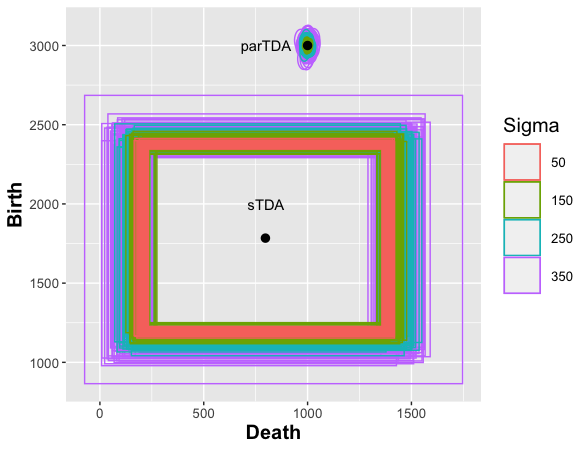}
        \caption{Confidence Regions} \label{subfig:CI}
  \end{subfigure}
\caption{death and birth estimates (A) and confidence regions (B) of 100 images across four noise levels, $\sigma=\{50, 150, 250, 350 \}$. (A) Point estimates for $(\mu_{1*}, \mu_1)$ using \textbf{tTDA} (triangle) and \textbf{parTDA} (asterisk), estimates of $(\tilde \mu_{1*}, \tilde \mu_1)$ using \textbf{sTDA} (plus), and the true death and birth time of the manifold (black circle). (B) The $95$\% confidence regions for $(\mu_{1*}, \mu_1)$  using \textbf{parTDA} and $(\tilde \mu_{1*}, \tilde \mu_1)$ using \textbf{sTDA}.} \label{fig:Simulations}
\end{figure}

Figure~\ref{fig:Simulations} illustrates the simulation results, with examples of point estimates for the death and birth times shown in Figure~\ref{subfig:estimates} (i.e., estimated pattern) and their corresponding confidence regions are shown in Figure~\ref{subfig:CI} (i.e., uncertainty estimate for the pattern). In both figures, each color represents a different $\sigma$ value. In Figure~\ref{subfig:estimates}, the shapes are the estimated death and birth times for each method where the black dots are the true death and birth time of the smoothed $(\tilde \mu_{1*},\tilde \mu_{1})$ and unsmoothed loop $(\mu_{1*}, \mu_{1})$. In Figure~\ref{subfig:CI}, the rectangles are the confidence regions using \textbf{sTDA} with the $L_\infty$ distance and the ellipses are the confidence regions generated using \textbf{parTDA}. These estimates and confidence regions from \textbf{parTDA} and \textbf{sTDA} are displayed together for convenience and to show the difference in scale, but recall that \textbf{parTDA} and \textbf{tTDA} are attempting to estimate the unsmoothed truth while \textbf{sTDA} is attempting to estimate the smoothed truth.

Across all noise settings, point estimates from \textbf{tTDA} in Figure~\ref{subfig:estimates} are significantly biased, especially as the noise level increases. While \textbf{parTDA} creates unbiased estimates close to $\mu_{1*}$ and $\mu_{1}$ and \textbf{sTDA} creates unbiased estimates close to $\tilde \mu_{1*}$ and $\tilde \mu_{1}$. However, the confidence regions created using \textbf{parTDA} are much smaller (more precise) compared to \textbf{sTDA}. Using \textbf{sTDA}, the confidence bands are large enough that a persistence of zero is within each confidence region for every loop in the data. This result suggests that no loop is distinctly identified within the underlying pattern. Whereas, \textbf{parTDA} correctly identifies one loop for all simulated images when using Algorithm~\ref{alg:Seg1H1}, and no other loops in the image are matched to the segmentation. In terms of coverage, \textbf{sTDA} covers the true death and birth times of $\mathcal{\tilde M}^0$ $100$\% of the time for a $95$\% confidence region. This conservative coverage is also present in checks of the original method using a KDE on point cloud data \cite{14} where we also found the empirical coverage to be $100$\%. In comparison, the coverage of the proposed \textbf{parTDA} method was always approximately $95$\% at all noise levels  considered. 
\begin{table}
\centering
\begin{tabular}{|>{\centering\arraybackslash}p{1.5cm}|
                    >{\centering\arraybackslash}p{1.2cm}|
                    >{\centering\arraybackslash}p{3.5cm}|
                    >{\centering\arraybackslash}p{3.2cm}|}
\hline
{\bf Method} & {\bf Noise Level} & {\bf Average confidence region area (SE)} & {\bf Average coverage percentage (SE)} \\
\hline
\textbf{sTDA} & 50 & 1390574 (5553.5)&  100\ (0)  \\
\cline{2-4}
     & 150 & 1460183  (9529.5) & 100 (0)   \\
\cline{2-4}
                 & 250 & 1577603  (14139.8)  & 100 (0)  \\
\cline{2-4}
                 & 350 & 1746974 (28184.4)  & 100 (0)  \\
\hline
\textbf{parTDA} & 50 & 122.9 (0.683) & 94.7 (0.2) \\
\cline{2-4}
                 & 150 & 1099.3 (7.732)  & 95.3 (0.2)  \\
\cline{2-4}
                 & 250 & 3057.2 (18.359) & 94.6 (0.2) \\
\cline{2-4}
                 & 350 & 5980.2(30.602) &  94.9 (0.3)  \\
\hline
\end{tabular}
\captionof{table}{Simulations results of a noisy loop for \textbf{sTDA} (rows 1-4) and \textbf{parTDA} (rows 5-8). The average confidence region area and standard errors (SE) are displayed for each noise level, based on 100 iid images in each setting. The fourth column is the percent coverage of the 95\% confidence regions, and corresponding SEs.} \label{tab:simulationResults}
\end{table}

\subsection{Runtime comparison} 

Computation of confidence regions for the death and birth times of topological features in a persistence diagram can be intensive. For instance, traditional TDA inference techniques often require repeated bootstrapping and smoothing of spaces (or repeated generation of persistence diagrams) followed by calculating the $L_{\infty}$ distance between these smoothed functions (or calculating the bottleneck distance between persistence diagrams). We evaluate the runtime of both methods using a simple example: a $32 \times 32$ image containing a single loop. The simulated data image is generated with an underlying structure where the true (death, birth) pair is ($\mu_{1*}=2000, \mu_1=3000$), a noise level of $\sigma=100$ for both partitions, and an adaptive smoothing parameter of 0.3 when implementing {\bf sTDA}. Table~\ref{tab:simulationTime} displays the computational times (in minutes) for both methods, computed on a MacBook Pro (Apple M3 Max, 64 GB RAM). The {\bf sTDA} method required approximately 4.06 minutes in total to estimate $(\tilde \mu_{1*},\tilde \mu_1)$ with  $(\tilde d_j, \tilde b_j)$ and construct confidence boxes. In contrast, the {\bf parTDA} method required approximately 0.14 minutes in total to estimate $(\mu_{1*}, \mu_1)$ with $(\bar Z_{1*}, \bar Z_1)$ and construct confidence ellipses. The extra preprocessing steps which are needed for {\bf parTDA}, such as the segmentation and the misclassification algorithm (Algorithm~\ref{alg:MisclassifiedPixels}), are included in the total time. In this simple example to calculate computational burden, the segmentation was implemented in R rather than Matlab to compare the computational times of the methods in the same environment.
    
    \begin{table}[h]
    \centering
    {\begin{tabular}{|c|c|c|c|c|}
    \hline
    {\bf Method} & {\bf Estimates} & {\bf Confidence Regions} & {\bf Segmentation} & \bf{Total} \\ \hline
    {\bf sTDA} & 0.02 & 4.04 & N/A & 4.06 \\ \hline
    {\bf parTDA}  & 0.03 & 0.039 & 0.07 & 0.14 \\ \hline
    \end{tabular}}
    \caption{Time results (in minutes) for applying {\bf sTDA} and {\bf parTDA} to an image with one loop where the segmentation includes segmenting the image and checking if any pixels are misclassified in the edge set using Algorithm~\ref{alg:MisclassifiedPixels}.}
    \label{tab:simulationTime}
    \end{table}

\subsection{Multiple loops simulation}
\begin{figure}
\centering
  \begin{subfigure}{.45\linewidth}
        \includegraphics[width=\linewidth]{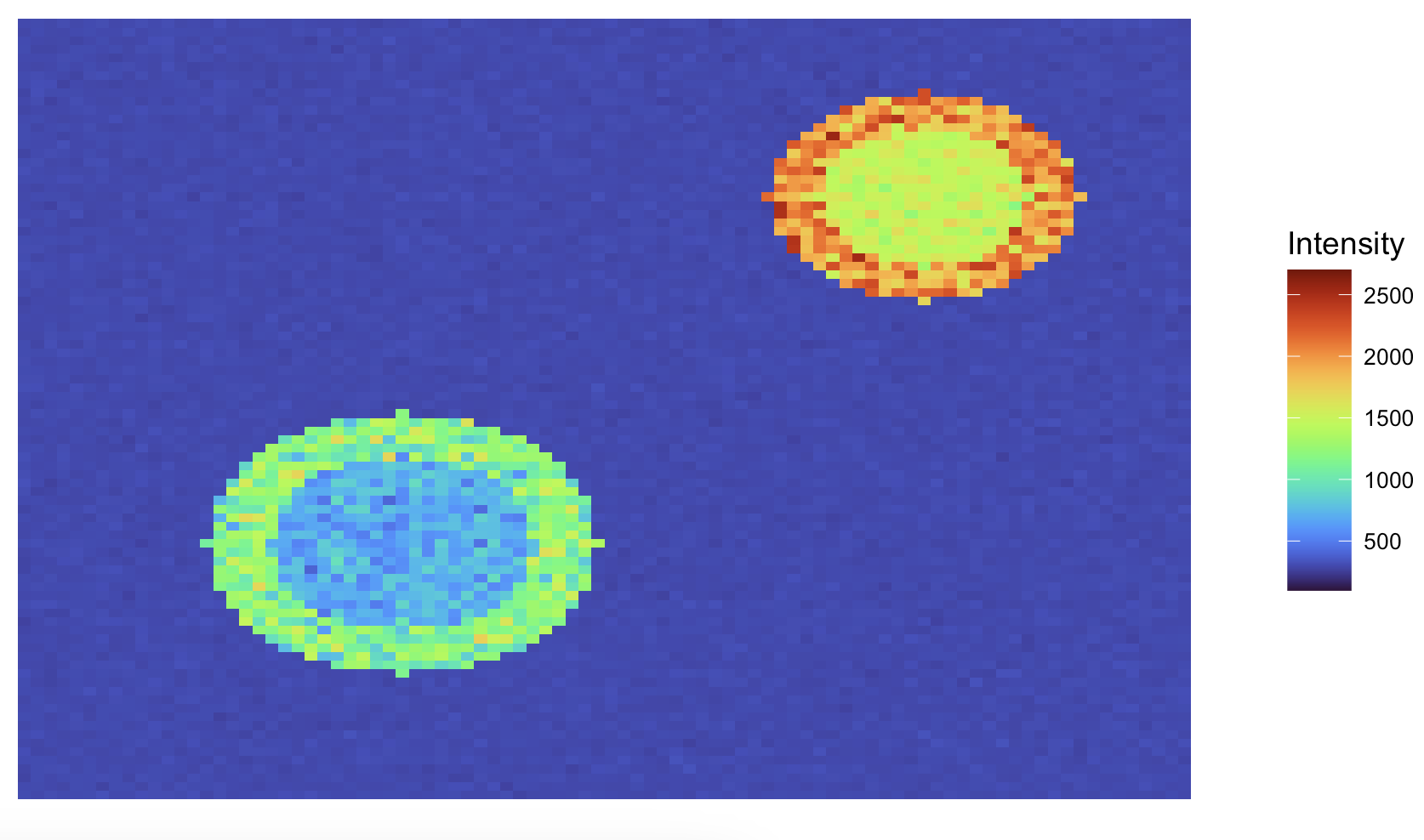}
         \caption{Image $\mathcal{M}$} \label{subfig:MLimage}
  \end{subfigure}  
  \begin{subfigure}{.4\linewidth}
    \centering
        \includegraphics[width=\linewidth]{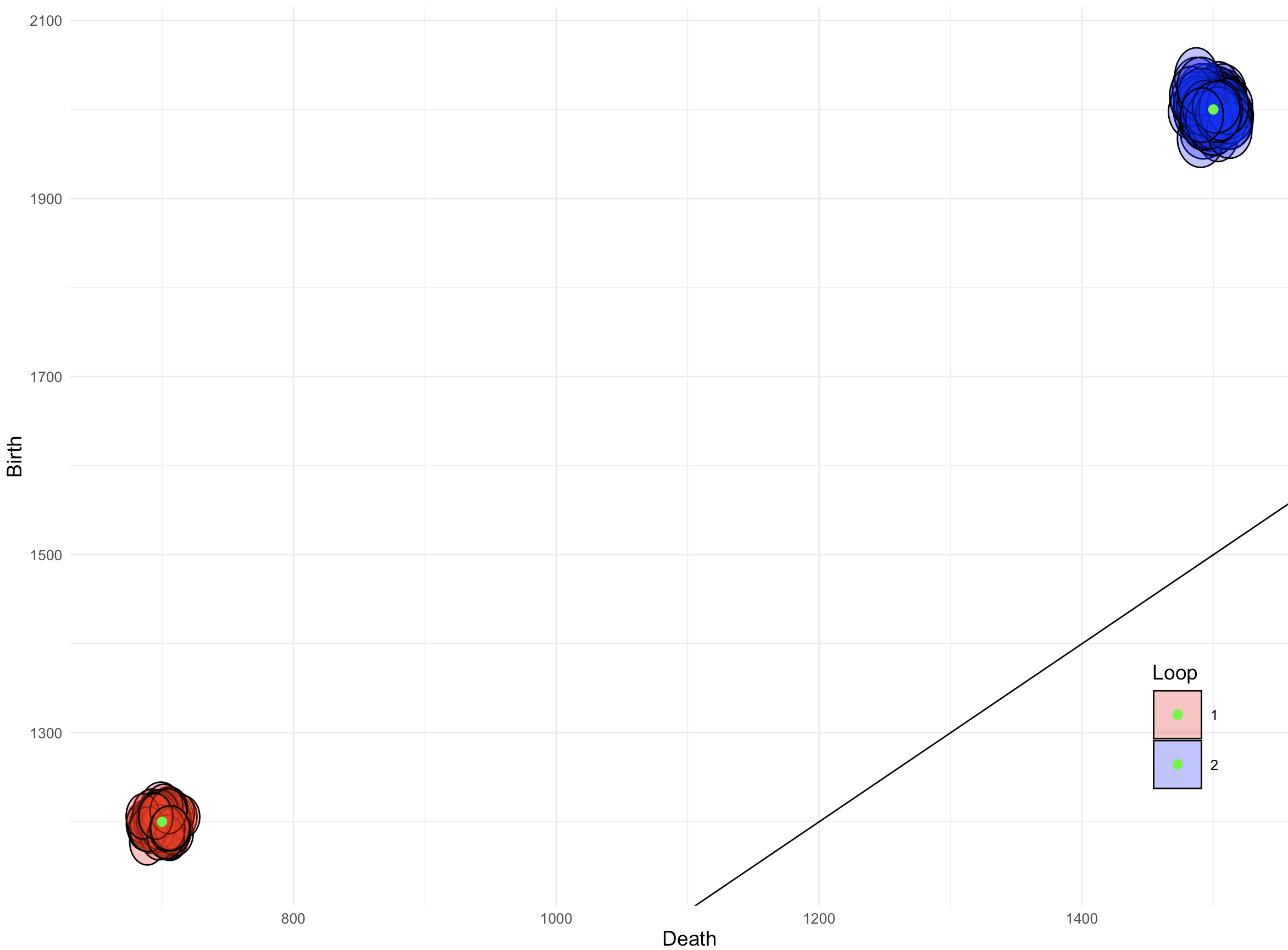}
         \caption{Confidence Regions}\label{subfig:MLCR}
  \end{subfigure}
\caption{(A) An example of simulated images with multiple loops. (B) The confidence regions for the first two most persistent loops (blue and red) where the green points are the death and birth times of the true loops.}\label{fig:MultLoopsEx}
\end{figure}

In order to evaluate the performance of {\bf parTDA} in the presence of multiple loops, $100$ images of dimension $90 \times 90$ were generated based on an underlying pattern containing two loops with death birth pairs $(\mu_{1*}, \mu_1) = (1500, 2000)$ and $(\mu_{2*}, \mu_2) = (700, 1200)$, respectively, and with background intensity $\mu_0=200$. As an illustrative example, Figure~\ref{subfig:MLimage} shows one of those $100$ generated image where the pixel values are sampled from normal distributions with noise levels of $\sigma_0=10$ for the background partition, $\sigma_{1*},\sigma_{2*}=100$ for the loop interior partitions, and $\sigma_1,\sigma_2=200$ for the loops. Assuming the true partitions are known (i.e., $\mathcal{G}_{1*},\mathcal{G}_1$ for loop 1 and $\mathcal{G}_{2*},\mathcal{G}_2$ for loop 2 are known), the empirical coverage of the 95\% confidence regions were computed after applying Algorithm~\ref{alg:Seg1H1} to match the death and birth times of loops in the persistence diagram of the image to the partitions of the loops. Then, with the known partitions matched to loops, Equation~\eqref{eq:CI} can be used to calculate the confidence ellipses for each loop separately. Figure~\ref{subfig:MLCR} shows the corresponding confidence ellipses from all 100 images where the true death and birth times are the green points and the color of each region is for the different loops.

When applying {\bf parTDA} to all $100$ images, the coverage of the 95\% confidence regions for loop 1 and loop 2 was $92\%$ and $94\%$, respectively. These values fall well within the expected range of coverage for a $95\%$ confidence region.

\section{Cell biology application}\label{sec:CellImages}

Pattern formation is a common and critically important feature of living systems. It is a natural process that occurs across biological scales ranging from ecosystems \cite{21, 22}, to developing tissues \cite{23, 3}, to individual cells \cite{25, 26}. Further, abnormal cell or tissue pattern formation is a feature of various pathological conditions, including cancers \cite{27, 28}. Consequently, approaches for objectively detecting and quantifying patterns and their quality are of interest for both basic biology and medicine. In this paper, pattern is assessed from the perspective of TDA through estimation of the death and birth times of rings with {\bf parTDA}. A higher persistence (birth-death) is indicative of stronger topological signal, and can be interpreted as a stronger pattern in this context.

\begin{figure}
\centering
  \begin{subfigure}{.49\linewidth}
    \centering
        \includegraphics[width=\linewidth]{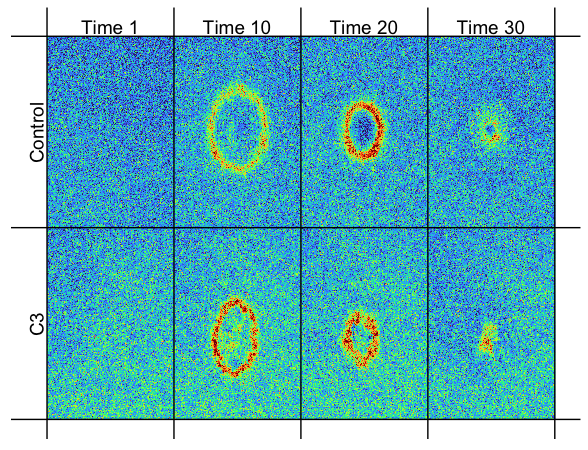}
        \caption{$\mathcal{M}^{\text{control}}_t$ and $\mathcal{M}^{\text{C3}}_t$ for $\{t_1,t_{10},t_{20},t_{30}\}$} \label{subfig:RealDataExample}
  \end{subfigure}
    \begin{subfigure}{.49\linewidth}
    \centering
        \includegraphics[width=\linewidth]{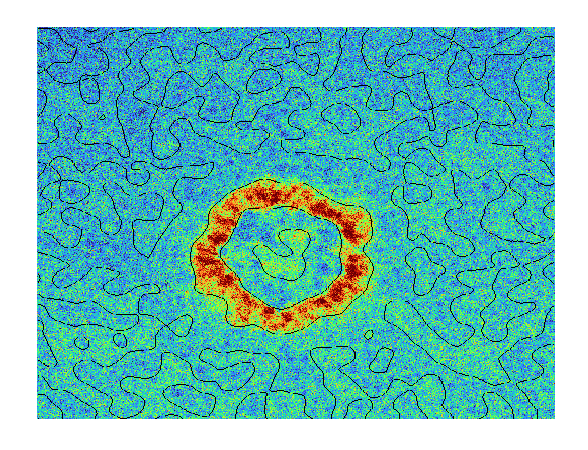}
        \caption{Segmentation $e^{\text{C3}}_{t_{15}}$} \label{subfig:SegmentationExample}
  \end{subfigure}
\caption{(A) The top row displays the images for the control cell $\mathcal{M}^{\text{control}}_t$ and the bottom row is the images for the C3 cell $\mathcal{M}^{\text{C3}}_t$. The columns represent different time points, $t_1$, $t_{10}$, $t_{20}$, and $t_{30}$. (B) Image of $\mathcal{M}^{\text{C3}}_{t_{15}}$ segmented by $e^{\text{C3}}_{15}$ where the black lines are the edges.} \label{fig:RealData}
\end{figure}

The proposed \textbf{parTDA} is applied to images of two individual cells sustaining wounds at distinct time points as illustrated in Figure~\ref{fig:RealData}. One of the cells was injected with a toxin (C3 exotransferase) that inhibits healing. The other cell is only wounded with no injection and serves as a control. The image for the C3 cell is denoted as $\mathcal{M}^{\text{C3}}_t$ and the image for the Control cell denoted as $\mathcal{M}^{\text{control}}_t$ for times $t=\{t_1, \ldots, t_{30}\}$. Time $t_1=0$ seconds is when the cell is wounded with sequential images separated by 8 seconds. Examples of the cell images at different time points are shown in Figure~\ref{subfig:RealDataExample}. Each of the images at every time point, $\mathcal{M}^{\text{control}}_t$ and $\mathcal{M}^{\text{C3}}_t$, was partitioned using the segmentation scheme from Section~\ref{sec:segmentation} with $e^{\text{control}}_t$ and $e^{\text{C3}}_t$ representing the edge sets at time $t$. An example of a segmentation at $t_{15}$ for $\mathcal{M}^{\text{C3}}_t$ is shown in Figure~\ref{subfig:SegmentationExample}.
The analysis is conducted independently at each time point. For each $t$ the number of rings in an image are detected using Algorithm~\ref{alg:Seg1H1} and a confidence region is created around the death and birth times using Equation~\eqref{eq:CI}. In this higher resolution image, Algorithm~\ref{alg:Seg1H1} has to be modified because multiple pixels in the image are equal to $b_j$. To address this, we smoothed the image, calculated the death and birth times, and used the smoothed birth time $\tilde b_j$ to help locate the pixel associated with $b_j$. 
 
For both $\mathcal{M}^{\text{C3}}$ and $\mathcal{M}^{\text{control}}$, no ring was detected until time $t_8$ for the {\bf parTDA} method even though the {\bf tTDA} method does detect rings in images for $t \leq t_7$. When using {\bf parTDA}, no ring was contained in $e^{\text{control}}_{t}$ and $e^{\text{C3}}_{t}$ for $t \leq t_7$, so Algorithm~\ref{alg:Seg1H1} has no partitions to match with the rings detected in {\bf tTDA}. From times $t_8$ to $t_{28}$ one ring is matched from $e^{\text{control}}_{t}$ to $\mathcal{M}^{\text{control}}_{t}$ and from $e^{\text{C3}}_{t}$ to $\mathcal{M}^{\text{C3}}_{t}$ using {\bf parTDA}, and thus, these are the times focused on in this section. 

\begin{figure}
\centering
  \begin{subfigure}{0.43\linewidth}
    \centering
        \includegraphics[width=\linewidth]{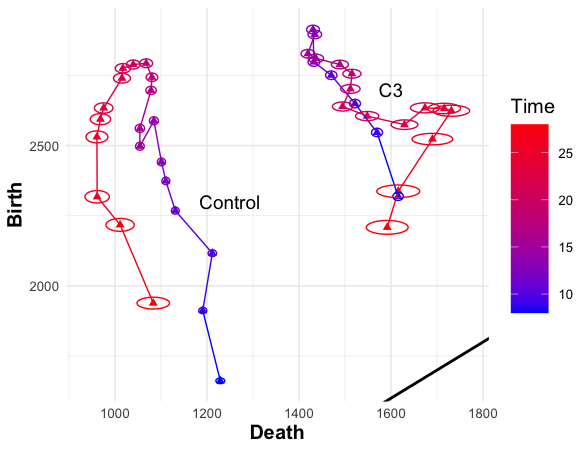}
        \caption{(Death, Birth) estimates} \label{subfig:bothPD}
  \end{subfigure}
    \begin{subfigure}{0.43\linewidth}
    \centering
        \includegraphics[width=\linewidth]{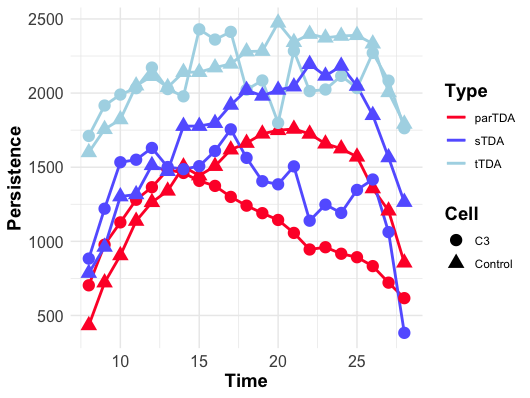}
        \caption{Estimated Persistence} \label{subfig:bothpersistence}
  \end{subfigure}
\caption{Estimated persistences of the C3 and Control cell images from $t=\{t_8,\ldots,t_{28}\}$. (A) The {\bf parTDA} death and birth times are shown on the persistence diagram along with confidence regions for both the C3 cell (right) and the Control cell (left). The black line is the diagonal line $b=d$. (B) Persistence is plotted over time for the C3 cell (solid line with points) and the Control cell (dashed line with triangles) using {\bf parTDA} (red), {\bf sTDA} (purple), and {\bf tTDA} (light blue).} \label{fig:RealDataResults}
\end{figure}

Two different visualizations of persistence across time for both cells are displayed in Figure~\ref{fig:RealDataResults}. In Figure~\ref{subfig:bothPD}, the {\bf parTDA} death and birth estimates are shown on a persistence diagram along with the confidence regions. The estimated death and birth times are connected by time, where time is indicated by different colors. Figure~\ref{subfig:bothpersistence}, is another way to visualize persistence (y-axis) over time (x-axis). When using {\bf parTDA}, the estimated persistence is $\bar Z^{1}_t-\bar Z^{1*}_t$, at each time $t$. The confidence set moves from a bivariate normal ellipse to a normal confidence interval centered at $\bar Z^{1}_t-\bar Z^{1*}_t$ with approximate variance $(\hat \sigma^2_{1})_t+(\hat \sigma^2_{1*})_t$. The red lines are the  estimated persistence and confidence intervals from {\bf parTDA} for both C3 (points) and Control (triangle) cells; the error bars are too small to see since sample size is large due to the high-resolution images. The dark blue lines use {\bf sTDA} and the light blue lines use {\bf tTDA} to estimate persistence across time; no confidence intervals were created for these methods. In general, {\bf sTDA} and {\bf tTDA} display more variability in the estimated persistences across time than {\bf parTDA}, and the C3 and Control cell persistences for {\bf tTDA} are not well separated. The overall trends in {\bf sTDA} and {\bf parTDA} are similar, though the {\bf parTDA} persistences appear to be more stable across time.

From $t_8$ to $t_{14}$, the most rapid growth in the persistence (or strength of pattern) are observed. Originally, the C3 cell images have more pattern in terms of the ring having a higher persistence than the Control cell images. However, at $t_{14}$ the wound ring in the Control cell continues to increase in its persistence while the wound ring in C3 cell begins to decline. In later time periods, the rings have shrunk in size, but not necessarily in intensity. The smaller size of the rings in the images result in larger confidence regions since the sample sizes of the sample means (i.e., the number of pixels in the pattern) has decreased. 
After $t_{29}$ the segmentation, $e^{\text{C3}}_{t_{29}}$, does not have any rings in the partitions; the edge set in the background is almost completely connected as one edge. Two distinct edges are needed to separate the section of an image into $\mathcal{M}^{\sigma}_i$ and $\mathcal{M}^{\sigma}_{i*}$ to find a ring in the segmentation. Therefore, no ring on $\mathcal{P}(\mathcal{M}^{\text{C3}}_{29})$ is matched to any regions in $e^{\text{C3}}_{t_{29}}$ as per Algorithm~\ref{alg:Seg1H1}.

During times $t_{29}-t_{30}$, the segmentation of the Control cell images continues to detect a ring where the wound is (i.e., two distinct edges separate $\mathcal{M}^{\text{Control}}_1$ and $\mathcal{M}^{\text{Control}}_{1*}$, which are matched to the ring detected in $\mathcal{P}(\mathcal{M}^{\text{Control}})$ for times $t_{29},t_{30}$); however, in order to directly compare the Control cell with the C3 cell only times $t_8-t_{28}$ are included in Figure~\ref{fig:RealDataResults}.

\section{Conclusions and discussions}
\label{sec:conc}
This paper includes three primary developments in TDA methodology. First, \textbf{parTDA} is proposed to estimate the death and birth times of topological features found in an image, which reduces the bias in the traditional TDA estimates (\textbf{tTDA}). Second, \textbf{parTDA} provides a process to quantify the uncertainty associated with these new death and birth time estimates in the form of a confidence region on a persistence diagram for an image. And finally, a persistence diagram confidence region method of \cite{14} was extended from point-cloud data to a single image as an alternative method (\textbf{sTDA}), which facilitated the creation of a new method to bootstrap an image. In general, \textbf{parTDA} is applicable to any image to determine the underlying pattern (in terms of holes) of that image and to quantify the uncertainty in that pattern. Though it was not necessary in our cell biology application, when comparing images with different pixel intensity ranges, it may be necessary to normalize the pixel intensities (e.g., dividing by the maximum pixel intensity in the image resulting in values from 0 to 1) so that detected differences are due to differences in topological structure rather than simply differences in the pixel intensity values.

Our novel {\bf parTDA} approach builds confidence regions on topological summary statistics (persistence diagrams) through estimating the mean and variance of the partitions associated with the death and birth times of homology group generators in the image. In this paper, we considered loops with uniform thickness due to the cell biology application. Irregularly shaped loops with varying thickness performs the same, as long as the partitioning of the image is done correctly and the method assumptions hold. These estimated means and variances use the Central Limit Theorem to get confidence ellipses for the death and birth times of loops in the persistence diagram. The sample means of pixels within the estimated death and birth time partitions of the manifold are represented on the persistence diagram through a matching procedure between {\bf parTDA} and {\bf tTDA} using Algorithm~\ref{alg:Seg1H1}. The {\bf parTDA} confidence regions are more accurate in terms of  coverage and have a smaller area than the alternative method, {\bf sTDA}.

The proposed methods were motivated by the goal of developing the means to objectively quantify patterns, especially loops, in a single image. As noted in the Introduction, loops are common features of biological samples and, in some cases such loops become distorted as a consequence of damage, disease, or other pathological insults such as toxins (e.g., \cite{5, 29}). While the differences in control and toxin-exposed ring patterns were subtle, the {\bf parTDA} method was nonetheless able to distinguish between the two, particularly at increasing times of healing. This finding bodes well for the application of {\bf parTDA} to other situations where biological ring organization is more obviously altered either due to deliberate experimental manipulations \cite{30} or due to disease \cite{29}. Further, independently of its ability to identify pattern differences within single examples of control and toxin-exposed samples, {\bf parTDA} was also able to distinguish between the control and toxin patterns when the entire time course of the experiment was taken into account such that there were no overlapping confidence regions. Confidence region comparisons are of particular utility to biologists and other scientists who seek to quantify the uncertainty in the ring structures in the images. The results presented here suggest that {\bf parTDA} may be useful beyond our initial goal of comparing individual patterns at a fixed time in that it may be possible to directly include time into the TDA analysis, through connecting loops across time and estimating the temporal uncertainty of the pattern. With these extensions, further investigation may be done to try and understand the mechanism at work when a cell is wounded under normal versus pathological conditions.

There are several extensions or improvements which could be made to {\bf parTDA}. First, {\bf parTDA} assumes independent and identically distributed pixel intensities on the loop (or any dimension $p$ topological feature of interest) which may not be the case in many applications. A generalization that may relax this assumption  could use localized averaging to estimate the birth and death time (instead of using the entire partition).

For future research, we aim to extend the {\bf parTDA} framework to include time and continuous functions or point-cloud data settings. An extension of {\bf parTDA} to point-cloud data can be directly compared to the methods of \cite{14}. In Section~\ref{sec:simulations}, {\bf sTDA} uses the $L_\infty$ distance between images to estimate the confidence regions as opposed to the bottleneck distance between persistence diagrams to be consistent with the \cite{14} approach. The confidence regions are smaller using the bottleneck distance (though still significantly larger than the {\bf parTDA} confidence regions), but the coverage is still at $100\%$. We are interested in investigating why these confidence regions are large for both the point cloud and image settings.

Another possible direction is to add a probabilistic element to the image segmentation, such as fuzzy clustering, to reduce false positive loops detected in the pattern. As mentioned in Section~\ref{sec:segmentation}, a limitation of parTDA is its reliance on correct segmentation. For instance, the segmentation may introduce a loop that is not part of the underlying pattern but is also matched to a loop found using {\bf tTDA}. While {\bf parTDA} is designed to build confidence regions, they can also be applicable to hypothesis testing to separate topological signal from noise. The performance of {\bf parTDA} as a hypothesis testing framework is a topic of future investigation.

\section*{Acknowledgments}
SG and JCK gratefully acknowledge support from NSF under Grant Number DMS 2038556. JCK gratefully acknowledges support from NSF under Grant Number 2337243. JZ gratefully acknowledges support from NSF under Grant Number DMS 2245906. WMB gratefully acknowledges support from NIH under Grant Number RO1 GM052932. Research presented in this article was supported by the National Security Education Center (NSEC) Informational Science and Technology Institute (ISTI) using the Laboratory Directed Research and Development program of Los Alamos National Laboratory under project number 20240479CR-IST.









\medskip
Received xxxx 20xx; revised xxxx 20xx; early access xxxx 20xx.
\medskip


\begin{thebibliography}{99}

\bibitem{22}
\newblock I. Barbier, H. Kusumawardhani, and Y. Schaerli,
\newblock \doititle{Engineering synthetic spatial patterns in microbial populations and communities},
\newblock \emph{Current Opinion in Microbiology}, 67 (2022): 102149.

\bibitem{26}
\newblock W. M. Bement, A. B. Goryachev, A. L. Miller, and G. von Dassow,
\newblock \doititle{Patterning of the cell cortex by Rho {GTPases}},
\newblock \emph{Nat Rev Mol Cell Biol}, 25.4 (2024): 290-308.

\bibitem{25}
\newblock W. M. Bement, A. L. Miller, G. von Dassow,
\newblock \doititle{Rho {GTPase} activity zones and transient contractile arrays},
\newblock \emph{Bioessays}, 28.10 (2006): 983-993.

\bibitem{10}
\newblock A. Bukkuri, N. Andor, and I. Darcy,
\newblock \doititle{Applications of Topological Data Analysis in Oncology},
\newblock \emph{Frontiers in Artificial Intelligence}, 4 (2021): 659037.

\bibitem{5}
\newblock B. M. Burkel, H. A. Benink, E. M. Vaughan, G. von Dassow, and W. M. Bement,
\newblock \doititle{A Rho {GTPase} signal treadmill backs a contractile array},
\newblock \emph{Developmental cell}, 23.2 (2012): 384-396.


\bibitem{18}
\newblock J. F. Canny,
\newblock \doititle{A Computational Approach to Edge Detection},
\newblock \emph{IEEE Transactions on Pattern Analysis and Machine Intelligence}, 6 (1986): 679-698.

\bibitem{15}
\newblock F. Chazal and B. Michel,
\newblock \doititle{An Introduction to Topological Data Analysis: Fundamental and Practical Aspects for Data Scientists},
\newblock \emph{Frontiers in Artificial Intelligence}, 4 (2021): 667963.

\bibitem{11}
\newblock M. K. Chung, P. Bubenik, and P. T. Kim.
\newblock \doititle{Persistence Diagrams of Cortical Surface Data},
\newblock \emph{International Conference on Information Processing in Medical Imaging}, Berlin, Heidelberg: Springer Berlin Heidelberg, 2009.

\bibitem{20}
\newblock D. Cohen-Steiner, H. Edelsbrunner, and J. Harer,
\newblock \doititle{Stability of Persistence Diagrams},
\newblock \emph{Proceedings of the twenty-first annual symposium on Computational geometry}, 2005.

\bibitem{dakurah2025maxtda}
\newblock S. Dakurah and J. Cisewski-Kehe,
\newblock \doititle{MaxTDA: Robust Statistical Inference for Maximal Persistence in Topological Data Analysis}
\newblock preprint, 2025, \arXiv{2504.03897}

\bibitem{16}
\newblock H. Edelsbrunner and J. Harer,
\newblock \emph{Computational Topology - an Introduction.}
\newblock American Mathematical Society, 2022.

\bibitem{14}
\newblock B. T. Fasy, F. Lecci, A. Rinaldo, L. Wasserman, S. Balakrishnan, A. Singh,
\newblock \doititle{Confidence sets for persistence diagrams},
\newblock \emph{The Annals of Statistics}, Institute of Mathematical Statistics., (2014): 2301-2339.

\bibitem{13}
\newblock S. Gupta, Y. Zhang, X. Hu, P. Prasanna, and C. Chen,
\newblock \doititle{Topology-Aware Uncertainty for Image Segmentation},
\newblock \emph{Advances in Neural Information Processing Systems,} 36 (2024).

\bibitem{2}
\newblock K. Haglund, I. P. Nezis, and H. Stenmark,
\newblock \doititle{Structure and functions of stable intercellular bridges formed by incomplete cytokinesis during development},
\newblock \emph{Communicative \& Integrative Biology,} 4.1 (2011): 1-9.

\bibitem{3}
\newblock J. C. Herron, S. Hu, B. Liu, T. Watanabe, K. M. Hahn, and T. C. Elston,
\newblock \doititle{Spatial models of pattern formation during phagocytosis},
\newblock \emph{PLOS Computational Biology}, 18.10 (2022).

\bibitem{29}
\newblock F. Le Naour, C. Sandt, C. Peng, N. Trcera, F. Chiappini, Flank, A. M., and Dumas, P. 
\newblock \doititle{In situ chemical composition analysis of cirrhosis by combining synchrotron fourier transform infrared and synchrotron X-ray fluorescence microspectroscopies on the same tissue section},
\newblock \emph{Analytical Chemistry}, 84 (2012):10260-10266

\bibitem{23}
\newblock A. Madamanchi, M.C. Mullins, and D. M. Umulis,
\newblock \doititle{Diversity and robustness of bone morphogenetic protein pattern formation},
\newblock \emph{Development}, 148.7 (2021): dev192344

\bibitem{4}
\newblock Craig A. Mandato and William M. Bement,
\newblock \doititle{Contraction and polymerization cooperate to assemble and close actomyosin rings around Xenopus oocyte wounds},
\newblock \emph{The Journal of Cell Biology}, 154.4 (2001): 785-798.

\bibitem{6}
\newblock Y. Mileyko, S. Mukherjee and J. Harer,
\newblock \doititle{Probability measures on the space of persistence diagrams},
\newblock \emph{Inverse Problems}, 27.12 (2011): 124007.

\bibitem{17}
\newblock N. Otter, M. A. Porter, U. Tillmann, P. Grindrod, and H. A. Harrington,
\newblock \doititle{A roadmap for the computation of persistent homology},
\newblock \emph{{EPJ} Data Science}, 6 (2017): 1-38.

\bibitem{27}
\newblock I. S. Paine and M. T. Lewis.
\newblock \doititle{The Terminal End Bud: the Little Engine that Could},
\newblock \emph{J Mammary Gland Biol Neoplasia}, 22 (2017): 93-108.

\bibitem{19}
\newblock J. R. Parker,
\newblock \emph{Algorithms for Image Processing and Computer Vision},
\newblock John Wiley \& Sons, 2010.

\bibitem{1}
\newblock T. D. Pollard and B. O'Shaughnessy,
\newblock \doititle{Molecular Mechanism of Cytokinesis},
\newblock \emph{Annual Review of Biochemistry}, 88.1 (2019): 661-689.

\bibitem{21}
\newblock R. M. Pringle and C. E. Tarnita,
\newblock \doititle{Spatial Self-Organization of Ecosystems: Integrating Multiple Mechanisms of Regular-Pattern Formation},
\newblock \emph{Annual Review of Entomology}, 62.1 (2017): 359-377.

\bibitem{30}
\newblock C. M. Simon,  E. M. Vaughan, W. M. Bement, and L. Edelstein-Keshet,
\newblock \doititle{Pattern formation of Rho GTPases in single cell wound healing.}
\newblock \emph{Molecular biology of the cell}, 24.3 (2013):421-432.

\bibitem{8}
\newblock Y. Singh, C. M. Farrelly, Q. A. Hathaway, T. Leiner, J. Jagtap, G. E. Carlsson, and B. J. Erickson,
\newblock \doititle{Topological data analysis in medical
imaging: current state of the art},
\newblock \emph{Insights Imaging}, 14.1 (2023): 58.

\bibitem{9}
\newblock Y. Skaf and R. Laubenbacher,
\newblock \doititle{Topological Data Analysis in Biomedicine: A Review},
\newblock \emph{Journal of Biomedical Informatics}, 130 (2022): 104082.

\bibitem{7}
\newblock K. Turner, Y. Mileyko, S. Mukherjee, and J. Harer,
\newblock \doititle{Fr{\'e}chet Means for Distributions of Persistence Diagrams},
\newblock \emph{Discrete \& Computational Geometry}, 52 (2014): 44-70.

\bibitem{28}
\newblock A. Uthamacumaran,
\newblock \doititle{Cancer: A turbulence problem},
\newblock \emph{Neoplasia}, 22.12 (2020): 759-769.

\bibitem{12}
\newblock J. Wang, K. Meng, F. Duan.
\newblock Hypothesis testing for medical imaging analysis via the smooth Euler characteristic transform.
\newblock preprint, 2023, \arXiv{2308.06645}.




\end{thebibliography}
\end{document}